\documentclass[aps,groupedaddress]{revtex4}
\usepackage{epsfig}
\usepackage{graphicx}
\usepackage[T1]{fontenc}
\usepackage{ae}
\usepackage{color}
\usepackage[latin1]{inputenc}
\usepackage{amssymb,amsbsy,amsmath}
\usepackage{bbm}

\newtheorem{prop}{Proposition}

\begin{document}



\title[Ground states]
{Classical ground states of spin lattices}

\author{Heinz-J\"urgen Schmidt$^1$ and Johannes Richter$^{2,3}$
}
\address{$^1$ Fachbereich Physik, Universit\"at Osnabr\"uck,
 D - 49069 Osnabr\"uck, Germany}
\address{$^2$ Institut f\"{u}r Physik, Otto-von-Guericke-Universit\"{a}t Magdeburg,
          P.O. Box 4120, D - 39016 Magdeburg, Germany}
\address{$^3$Max-Planck-Institut f\"{u}r Physik komplexer Systeme,
          N\"{o}thnitzer Stra\ss e 38, D - 01187 Dresden, Germany}


\begin{abstract}
We present a generalization of the Luttinger-Tisza-Lyons-Kaplan (LTLK) theory of classical ground states of Bravais lattices
with Heisenberg coupling to non-Bravais lattices.
It consists of adding certain Lagrange parameters to the diagonal of the Fourier transformed coupling matrix analogous to the
theory of the general ground state problem already published.
This approach is illustrated by an application to a modified honeycomb lattice,
which has exclusive three-dimensional ground states
as well as a classical spin-liquid ground state
for different values of the two coupling constants.
Another example, the modified square lattice,
shows that we can also obtain so-called incommensurable ground states by our method.
\end{abstract}

\maketitle

\section{Introduction}\label{sec:I}

A fundamental property of spin systems including lattices is the set of ground states and the ground state energy per site.
The classical limit of these quantities is also of interest and has been studied in numerous papers, of which we will only
mention a selection here \cite{LT46,LK60,L74,FF74,V77,SL03,N04,KM07,LH12,XW13,S17a,Getal20}.
Classical ground states are not only relevant for spin systems with large
spin but often define the relevant order parameter in quantum systems.
Moreover, they serve as starting point for quantum methods such as the spin wave
theory and the coupled cluster method, see, e.~g., \cite{C92,ZDR09,Getal11,CZ14}.
To detect the classical ground state of spin systems can be difficult in
presence of frustration, where non-collinear spin orders, incommensurate spiral phases,
or massively degenerated ground state manifolds may appear, see, e.~g., \cite{RTR79,MLM11}.
An additional  challenge appears if the lattice primitive unit cell contains more than one
site and/or the exchange couplings extend beyond the nearest-neighbor
separation.

The theory of classical ground states of spin lattices can be traced back to the seminal paper of J.~M.~Luttinger  and L.~Tisza \cite{LT46}
that is, however confined to dipole interaction. The Luttinger-Tisza approach has been generalized to Heisenberg spin systems
with general coupling coefficients $J_{\mu\nu}$ by
D.~H.~Lyons and T.~A.~Kaplan \cite{LK60,KM07} and can be summarized as follows: The problem of finding the minimum of
$\sum_{\mu\nu}J_{\mu\nu }{\mathbf s}_\mu\cdot{\mathbf s}_\nu$ subject to the ``strong constraint"
${\mathbf s}_\mu\cdot{\mathbf s}_\mu=1$ for all $\mu=1,\ldots,M$ can be replaced by the same problem, but with the ``weak constraint"
$\sum_\mu{\mathbf s}_\mu\cdot{\mathbf s}_\mu=M$, where $M$ denotes the number of spins (see Section \ref{sec:GS} for the detailed definitions).
The latter problem is solved by finding the minimal eigenvalue of the symmetric $J$-matrix with entries $J_{\mu\nu }$.
This eigenvalue problem is simplified by accounting for the invariance of the $J$-matrix under lattice translation
which leads to the consideration of the Fourier transformed $J$-matrix $\widehat{J}({\mathbf q})$.
It may happen that the ground state constructed by superpositions of the eigenvectors corresponding to the minimal eigenvalue already satisfies the
strong constraint. This will be the case for Bravais lattices, as shown in  \cite{LK60}, but may also hold for some non-Bravais
lattices. The problem remains to solve the ground state problem in cases where the LTLK theory
does not work since the strong constraint cannot be replaced by the weak constraint.

Our approach will be based on the ``Lagrange-variety theory" of classical ground states given in \cite{SL03,S17a,S17b,S17c,S17d,SF20a,SF20b}
which is intended to apply to general Heisenberg systems and is not specifically tailored for spin lattices.
The basic result of this theory is that the ground states can be constructed
by superpositions of the eigenvectors corresponding to the minimal eigenvalue
{\em not} of the original $J$-matrix but of the ``dressed $J$-matrix". The latter is obtained by adding certain Lagrange parameters
to the diagonal of the $J$-matrix. In principle, this approach is already mentioned in \cite{KM07},
but is considered too complicated, since one would have to find a macroscopic number of $10^{23}$ or so of Lagrangian parameters.
However, as mentioned in \cite{S17a}, the presence of symmetries reduces the number of independent Lagrange parameters.
In the case of lattice translation symmetry this means that we only have $L-1$ unknown parameters if $L$ is the number of Bravais
lattices needed to construct the spin lattice under consideration. This makes it plausible that our approach reduces to the
LTLK approach in the case of a Bravais lattice, where $L=1$.
It is worth mentioning that our approach is able to detect non-coplanar (i.e.
3-dimensional) ground states which may lead to the emergence of scalar chiral
orders as discussed for the celebrated highly frustrated kagome magnet,
see Refs.~\cite{Detal05,JRR08,MLM11}.

The paper is organized as follows. In Section \ref{sec:GS} we recapitulate the general definitions and results that have already appeared in
\cite{SL03,S17a,S17b,S17c,S17d,SF20a,SF20b}, while the specializations to spin lattices are presented in Section \ref{sec:SL}.
The proofs of two propositions appearing in this Section are moved to the Appendix.
For a first illustration, we consider the cyclic sawtooth chain consisting of $6$ corner-sharing triangles. The coplanar
ground states that can be also obtained by elementary considerations are shown to result from our approach in the Sections \ref{sec:SM}
and \ref{sec:OC}. In Section \ref{sec:TG} it is shown how to obtain a three-dimensional, less elementary ground state.
Some remarks on infinite lattices and incommensurable ground states are made in Section \ref{sec:IL}.
The main applications are contained in Section
\ref{sec:AP}.
After some general remarks in subsection \ref{sec:GR} on other possible cases
we concentrate, in subsection \ref{sec:AH}, on the honeycomb lattice with two different coupling constants
written as $J_1=\sin\phi$ and $J_2=\cos\phi$. where $-\pi< \phi\le\pi$.
In the sectors $|\phi|\ge \pi/2$ and slightly beyond we have one-dimensional ground states that can be understood elementarily.
The crucial application is the construction of two different phases of three-dimensional ground states in the sectors
$|\phi|< \pi/4$ and $\pi/4 <|\phi| < \arctan(3)$ and the corresponding closed formulas for the ground state energy.
The energies of all three mentioned phases are shown to be assumed by ground states calculated numerically.

Subsection \ref{sec:MSL} contains an example, the modified square lattice, where incommensurable ground states occur.
Although this problem is, strictly speaking, beyond the present theory, we will consider an extrapolation of our method
that leads to a semi-analytical calculation of the incommensurable ground states that have a lower energy than the ground states
of a finite model that can be numerically calculated.
In Section \ref{sec:SUM}, we summarize our study in a recipe-like manner
and discuss the question of whether this finds all ground states.

Throughout this paper we will call spin configurations either ``three-dimensional" (non-coplanar) or ``two-dimensional"
(coplanar) or ``one-dimensional" (collinear, Ising states), depending on the dimension of the linear space spanned by all spin vectors,
and avoid the other names given in brackets.

\section{Definitions and results}\label{sec:DR}

\subsection{General spin systems}\label{sec:GS}

We consider a finite system ${\mathbf s}$ of $M$ classical spin vectors ${\mathbf s}_\mu,\,\mu\in{\mathcal M},$ of unit length,
i.~e., satisfying
\begin{equation}\label{D2}
 {\mathbf s}_\mu \cdot {\mathbf s}_\mu=1 \quad \mbox{ for all } \mu\in{\mathcal M}
 \;.
\end{equation}
It will be convenient to consider ${\mathbf s}$ as an $M\times K$-matrix consisting of $M$ rows ${\mathbf s}_\mu$
and $K$ columns ${\mathbf s}^{(i)}$.
{The dimension $\text{dim}$ of the spin configuration is defined as the rank of the matrix ${\mathbf s}$
and is restricted to $\text{dim}=1,2,3$ for physical cases.}
The spin system will be called a ``spin lattice" if the index set ${\mathcal M}$ has the structure
\begin{equation}\label{structureM}
 {\mathcal M} = {\mathcal L}\times {\mathbbm Z}_{\mathbf N}:=  {\mathcal L}\times {\mathbbm Z}_{N_1}\times \ldots \times {\mathbbm Z}_{N_d}
 \;,
\end{equation}
where ${\mathcal L}$ is a finite set of size $|{\mathcal L}|=L$
(``primitive unit cell").
The finite model will consist of $\prod_{i=1,\ldots,d}N_i$ copies of the primitive unit cell
and ${\mathbbm Z}_{N_i}$ denotes the range of integers modulo $N_i$ (cyclic boundary conditions).
$d$ is the dimension of the lattice and will be restricted to the physical cases of $d=1,2,3$.
The geometry of the lattice in real space will not play any role here; in particular,
the distinction between the $14$ Bravais lattices for $L=1$ and $d=3$ does not matter.
This may be different in concrete applications when additional symmetries beyond the translational ones become important.

We assume a Heisenberg Hamiltonian of the form
\begin{equation}\label{HeisHam}
 H({\mathbf s})=\sum_{\mu,\nu\in{\mathcal M}}J_{\mu\nu}\,{\mathbf s}_\mu\cdot {\mathbf s}_\nu
 \;,
\end{equation}
with real coupling coefficients $J_{\mu\nu}$ satisfying $J_{\mu\nu}=J_{\nu\mu}$ and $J_{\mu\mu}=0$ for all $\mu,\nu\in{\mathcal M}$.
These coefficients can hence be viewed as the entries of a symmetric $M\times M$-matrix ${\mathbbm J}$.
The Hamiltonian (\ref{HeisHam}) does not uniquely determine ${\mathbbm J}$:
Let $\lambda_\mu,\,\mu=1,\ldots,M$
be arbitrary real numbers subject to the constraint
\begin{equation}\label{D4}
\sum_{\mu=1}^M \lambda_\mu =0
\;,
\end{equation}
and define a new matrix ${\mathbbm J}({\boldsymbol\lambda})$ with entries
\begin{equation}\label{D5}
J({\boldsymbol\lambda})_{\mu\nu}:= J_{\mu\nu}+\delta_{\mu\nu}\lambda_\mu
\;,
\end{equation}
then
\begin{eqnarray}\label{D6a}
\tilde{H}({\mathbf s})&:=& \sum_{\mu,\nu=1}^M J({\boldsymbol\lambda})_{\mu \nu}\,\mathbf{s}_\mu\cdot \mathbf{s}_\nu\\
\label{D6b}
&\stackrel{(\ref{D5})}{=}&
\sum_{\mu,\nu=1}^M J_{\mu \nu}\,\mathbf{s}_\mu\cdot \mathbf{s}_\nu+
\sum_{\mu=1}^M \lambda_\mu\,\mathbf{s}_\mu\cdot \mathbf{s}_\mu\\
\label{D6c}
&=& H({\mathbf s})
\;,
\end{eqnarray}
due to (\ref{D2}) and (\ref{D4}). The transformation
$J_{\mu\nu} \rightarrow J({\boldsymbol\lambda})_{\mu\nu}$ according to (\ref{D5}) has been called a ``gauge transformation"
in \cite{SL03} according to the close analogy with other branches of physics where this notion is common.
Thus, the Heisenberg Hamiltonian does not depend on the gauge.
In most problems the simplest gauge would be the ``zero gauge", i.~e.~, setting $\lambda_\mu=0$ for $\mu=1,\ldots,M$.
However, in the present context it is crucial not to remove the gauge freedom by a certain choice of the $\lambda_\mu$
but to retain it. We will hence explicitly stress the dependence of the coupling matrix on the undetermined
$\lambda_\mu$ by using the notation ${\mathbbm J}({\boldsymbol\lambda})$. ${\mathbbm J}({\boldsymbol\lambda})$ will be
called the ``dressed ${\mathbbm J}$-matrix" and its entries will be, as above, denoted by $J({\boldsymbol\lambda})_{\mu\nu}$.
The rationale is that we want to trace back
the properties of ground states to the eigenvalues and eigenvectors of
${\mathbbm J}({\boldsymbol\lambda})$ and these in a non-trivial way depend on ${\boldsymbol\lambda}$.
The ``undressed" matrix ${\mathbbm J}$ without ${\boldsymbol\lambda}$ will always denote a symmetric $M\times M$-matrix in the zero gauge.
Let $\Lambda$ denote the $M-1$-dimensional subspace of ${\mathbb R}^M$  defined by
\begin{equation}\label{DLambda}
\Lambda:= \left\{{\boldsymbol\lambda}\in{\mathbb R}^M\left| \sum_{\mu=1}^M\,\lambda_\mu=0\right.\right\}
\end{equation}
As coordinates in $\Lambda$ we will use the first $M-1$ components $\lambda_i,\,i=1,\ldots,M$ since
the $M$-th component can be expressed by the others via $\lambda_M=-\sum_{i=1}^{M-1} \lambda_i$.

Let us, for arbitrary ${\boldsymbol\lambda}\in\Lambda$, denote by $\jmath_{\scriptsize min}(\boldsymbol\lambda)$
the lowest eigenvalue of the real, symmetric matrix ${\mathbbm J}({\boldsymbol\lambda})$. Then by the Ritz-Rayleigh
variational priciple
\begin{eqnarray}
\label{Ritz1}
  H({\mathbf s}) &\stackrel{(\ref{D6a}-\ref{D6c})}{=}&\sum_{\mu,\nu=1}^M J({\boldsymbol\lambda})_{\mu \nu}\,\mathbf{s}_\mu\cdot \mathbf{s}_\nu \\
  \label{Ritz2}
   &\ge& \jmath_{\scriptsize min}(\boldsymbol\lambda) \sum_{\mu=1}^{M}\mathbf{s}_\mu\cdot \mathbf{s}_\mu \\
   \label{Ritz3}
   &\stackrel{(\ref{D2})}{=}& M\, \jmath_{\scriptsize min}(\boldsymbol\lambda)
   \;.
\end{eqnarray}
This holds also for ground states $\hat{\mathbf s}$ such that $H(\hat{\mathbf s})=E_{\scriptsize min}$. Hence for
arbitrary ${\boldsymbol\lambda}\in\Lambda$ we obtain the upper bound
\begin{equation}\label{upper}
  \jmath_{\scriptsize min}(\boldsymbol\lambda)\le \frac{1}{M}\,E_{\scriptsize min}
  \;,
\end{equation}
that will play a role in the definition of the critical point in (\ref{ua}).
The inequality (\ref{upper}) holds for all gauges $(\boldsymbol\lambda)$, especially for $(\boldsymbol\lambda)={\mathbf 0}$.
The latter will be called the ``Luttinger-Tizsa lower bound" to the minimal energy (per spin site).
In the present paper a special gauge will become important that we will call the ``ground state gauge".
It is well-known that a smooth function of a finite number of variables has a vanishing gradient at those points where it assumes its
(local or global) minimum.
If the definition domain of the function is constrained, as in our case of the function ${\mathbf s}\to H({\mathbf s})$,
its gradient no longer vanishes at the minima but will only be perpendicular
to the ``constraint manifold".
The resulting  equation reads, in our case,
\begin{equation}\label{D7}
\sum_{\nu=1}^M J_{\mu\nu}{\mathbf s}_\nu = - \kappa_\mu\,{\mathbf s}_\mu,\quad \mu=1,\ldots,M
\;.
\end{equation}
Here the $\kappa_\mu$  are the Lagrange parameters due to the constraints (\ref{D2}). This equation is only
necessary but not sufficient for ${\mathbf s}$ being a ground state. If it is satisfied we call the corresponding state a "stationary state" and will
refer to (\ref{D7}) as the ``stationary state equation" (SSE). This wording of course reflects the fact that exactly the stationary states
will not move according to the equation of motion for classical spin systems, see, e.~g., \cite{SL03}, but we will not dwell upon this here.
All ground states are stationary states but there are stationary states that are not ground states.
Let us rewrite (\ref{D7}) in the following way:
\begin{equation}\label{D8}
\sum_{\nu=1}^M J_{\mu\nu}{\mathbf s}_\nu = (\bar{\kappa}- \kappa_\mu)\,{\mathbf s}_\mu-\bar{\kappa}\,{\mathbf s}_\mu
=-\lambda_\mu\,{\mathbf s}_\mu-\bar{\kappa}\,{\mathbf s}_\mu
\;,
\end{equation}
where we have introduced the mean value of the Lagrange parameters
\begin{equation}\label{D9}
\bar{\kappa}:= \frac{1}{M}\sum_{\mu=1}^M\,\kappa_\mu
\;,
\end{equation}
and the deviations from the mean value
\begin{equation}\label{D10}
\lambda_\mu:= \kappa_\mu-\bar{\kappa},\;\mu=1,\ldots,M
\;.
\end{equation}

We denote by $\Lambda_0\subset\Lambda$ the set of vectors
${\boldsymbol\lambda}$ with components (\ref{D10}) resulting from
(\ref{D7}) in the case of a ground state $\hat{\mathbf s}$ with unrestricted dimension $K$. It has been
proven \cite{S17a} that $\Lambda_0$ consists of a single point $\Lambda_0=\{\hat{\boldsymbol\lambda}\}$
where the upper bound (\ref{upper}) is assumed:
\begin{equation}\label{ua}
 \jmath_{\scriptsize min}(\hat{\boldsymbol\lambda}) = \frac{1}{M}\,E_{\scriptsize min}
 \;.
\end{equation}
We will refer to this point as the ``critical point" of the function
${\boldsymbol\lambda}\mapsto \jmath_{\scriptsize min}({\boldsymbol\lambda})$.

${\boldsymbol\lambda}=\hat{\boldsymbol\lambda}$ will be called a ``ground state gauge".
It can be used for a gauge transformation $J_{\mu\nu}
\rightarrow J(\hat{\boldsymbol\lambda})_{\mu\nu}$ which renders (\ref{D8}) in the
form of an eigenvalue equation:
\begin{eqnarray}\label{D11a}
\sum_{\nu=1}^M J({\hat{\boldsymbol\lambda}})_{\mu\nu}\hat{\mathbf s}_{\nu}& =&  -\bar{\kappa}\,\hat{\mathbf s}_{\mu}\\
\label{D11b}
&=&\jmath_{\scriptsize min}(\hat{\boldsymbol\lambda})\,\hat{\mathbf s}_\mu
\;,
\end{eqnarray}
where the identification $\jmath_{\scriptsize min}(\hat{\boldsymbol\lambda})=-\bar{\kappa}$ follows from (\ref{ua}).
This equation can be written in matrix form if we recall that $\hat{\mathbf s}$ can be viewed as a matrix consisting of
$M$ rows $\hat{\mathbf s}_\mu$ and $K$ columns:
\begin{equation}\label{D12}
J({\hat{\boldsymbol\lambda}})\,\hat{\mathbf s}=\jmath_{\scriptsize min}(\hat{\boldsymbol\lambda})\,\hat{\mathbf s}
\;.
\end{equation}
This means that each column $\hat{\mathbf s}^{(i)},\; i=1,\ldots,K,$ of the matrix $\hat{\mathbf s}$ will be an eigenvector of
${\mathbbm J}(\hat{\boldsymbol\lambda})$ corresponding to the eigenvalue $\jmath_{\scriptsize min}(\hat{\boldsymbol\lambda})$.

Summarizing the theory developed so far, we can find the ground state spin configuration $\hat{\mathbf s}$ as a
suitable superposition
of eigenvectors corresponding to the minimal eigenvalue $\jmath_{\scriptsize min}(\hat{\boldsymbol\lambda})$ of the
dressed $J$-matrix ${\mathbbm J}({\boldsymbol\lambda})$ at the critical point $\hat{\boldsymbol\lambda}$.

{We emphasize that while the critical point $\hat{\boldsymbol\lambda}$ is unique,
the ground-state spin configuration $\hat{\mathbf s}$ is not.
Besides the ``trivial'' degeneracy due to rotations or reflections, there may be other ``additional'' degeneracies.
Recall that the dimension of the ground state was left open.
Therefore, it is even possible that different ground states have different dimensions.
In this case, the minimal dimension of the set of ground states is an important quantity.
In \cite{S17a} an example is given for a system with $N=10$ spins, where the minimal dimension
of the set of ground states is $4$ with ground state energy $E_\text{min}=-60$.
In such a situation, one would have to search for physical states,
i.e., with maximal dimension $3$, which realize the minimum energy $E_\text{min}^\text{phys}$ under the dimensional constraint,
which is $E_\text{min}^\text{phys}=-�59.1728$ in the mentioned example.
The present theory is not particularly suitable to solve this problem.}

\subsection{Spin lattices}\label{sec:SL}

The preceding considerations are quite general and do not use the lattice structure.
We will now consider this and write the indices $\mu\in{\mathcal M}$ as $\mu=(i,{\mathbf n})$
where $i\in{\mathcal L}$ and ${\mathbf n}\in {\mathbbm Z}_{\mathbf N}$.
In the language of solid-state physics the (multi)indices  ${\mathbf n}$ and  ${\mathbf m}$
label unit cells and $i$ and $j$  sites in a unit cell.
The coupling coefficients are
correspondingly written in the form $J_{\mu\nu}=J_{{\mathbf n}{\mathbf m}}^{ij}$ such that the symmetry requirement reads
\begin{equation}\label{symmJ}
J_{{\mathbf n}{\mathbf m}}^{ij}=J_{{\mathbf m}{\mathbf n}}^{ji}
\;,
\end{equation}
for all $i,j\in{\mathcal L}$ and ${\mathbf n},{\mathbf m}\in {\mathbbm Z}_{\mathbf N}$.
The crucial assumption throughout this paper is the invariance of the coupling under lattice translations:
\begin{equation}\label{latticetrans}
 J_{{\mathbf n}+{\mathbf a},{\mathbf m}+{\mathbf a}}^{ij}=J_{{\mathbf m},{\mathbf n}}^{ij}
 \;,
\end{equation}
for all ${\mathbf a}\in {\mathbbm Z}_{\mathbf N}$, where the addition ${\mathbf n}+{\mathbf a}$ is understood
modulo $N_i$ for the $i-$th component of the $d$-dimensional vector ${\mathbf n}+{\mathbf a}$.\\

\begin{figure}[htp]
\centering
\includegraphics[width=1.0\linewidth]{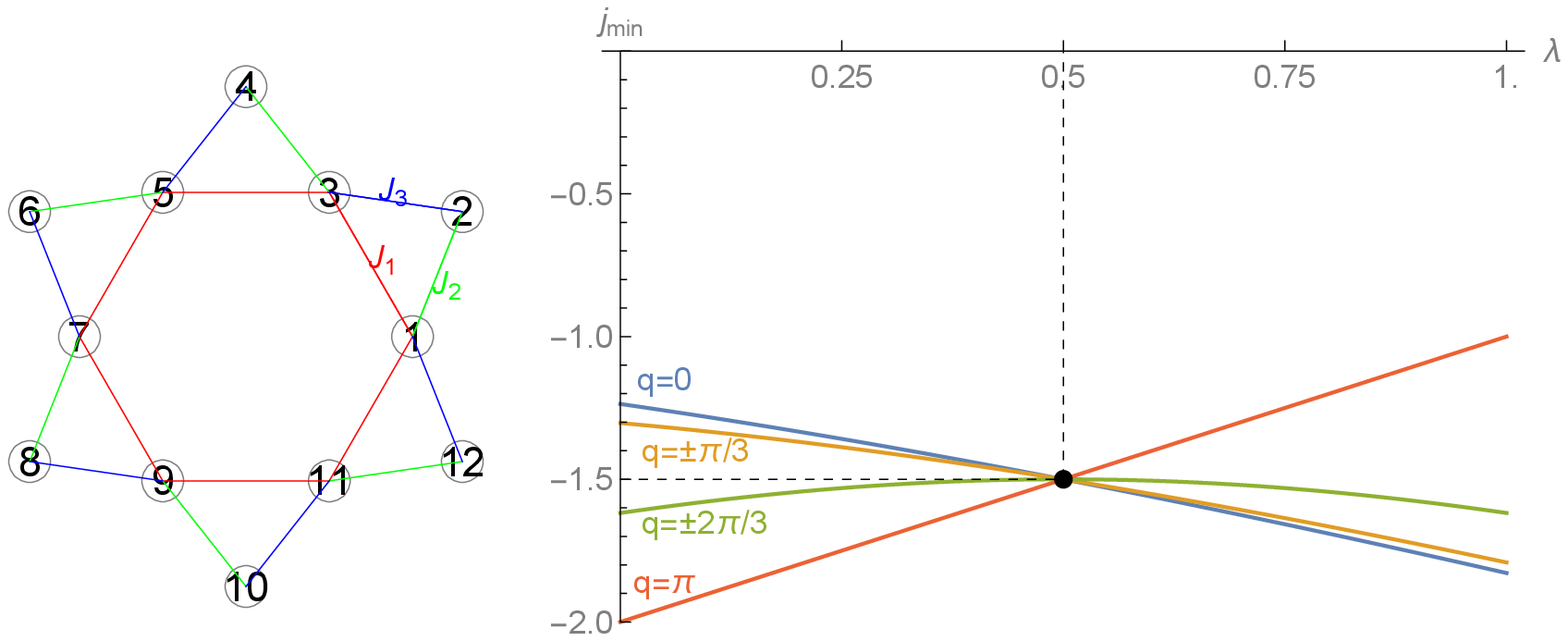}
\caption{{Left panel: Plot of the cyclic sawtooth chain consisting of $M=12$ spins.
The three coupling constants denoted by $J_1, \,J_2$ and $J_3$
are invariant under $60^\circ$ rotations.}\\
{Right panel: Plot of  the four functions $\lambda\mapsto \jmath_{\scriptsize min}(\lambda,q)$ corresponding
to $q=0,\pm\frac{\pi}{3},\pm\frac{2\pi}{3},\pi$ and the cyclic sawtooth chain with $J_1=J_2=J_3=1$.
The four graphs meet at the critical point (black dot) with coordinates $\hat{\lambda}=1/2$ and
$\jmath_{\scriptsize min}(\hat{\lambda},q)=-3/2$.
At this point the minimum of the four functions assumes its maximum according to the ground state gauge
condition (\ref{groundgauge})}.
}
\label{FIGCP}
\end{figure}

As a simple standard example we consider the ``cyclic sawtooth chain", see Figure \ref{FIGCP},
{left panel,} consisting of $M=12$ spin sites
arranged at $L=2$ concentric hexagons. This is a  $d=1$ dimensional lattice with index set ${\mathbbm Z}_6$. There are three different
couplings $J_1,\,J_2$ and $J_3$ between adjacent sites that are, according to (\ref{latticetrans}),  invariant under lattice translations, i.~e.,
under $60^\circ$ rotations.\\

It has been shown \cite{S17a} that the ground state gauge $\hat{\boldsymbol\lambda}$ has the same symmetries as the undressed $J$-matrix.
In our case this means that $\hat{\lambda}_{i,{\mathbf n}}$ only depends on $i\in{\mathcal L}$ and thus, slightly changing the notation,
the dressed ${\mathbbm J}$-matrix assumes the form
\begin{equation}\label{dressedJ}
 {\mathbbm J}_{{\mathbf n} {\mathbf m}}^{ij}(\hat{\boldsymbol\lambda})=J_{{\mathbf n}{\mathbf m}}^{ij}+\hat{\lambda}_i\,\delta_{ij}
 \;,
\end{equation}
again satisfying
\begin{equation}\label{sumlambda}
 \sum_{i=1}^{L}\hat{\lambda}_i=0
 \;.
\end{equation}

The eigenvalue equation (\ref{D11b}) following from the SSE (\ref{D7}) then assumes the form
\begin{equation}\label{eiglatt}
\sum_{j{\mathbf m}}{\mathbbm J}_{{\mathbf n}{\mathbf m}}^{ij}(\hat{\boldsymbol\lambda})\,\hat{\mathbf s}_{j{\mathbf m}}=
\jmath_{\scriptsize min}(\hat{\boldsymbol\lambda})\,\hat{\mathbf s}_{i{\mathbf n}}
\;.
\end{equation}
As mentioned above the ground state gauge $\hat{\boldsymbol\lambda}$ is uniquely determined as the point where the
function ${\boldsymbol\lambda}\mapsto \jmath_{\scriptsize min}({\boldsymbol\lambda})$ assumes a global maximum. For large systems it will be
difficult to calculate this ground state gauge and the corresponding eigenvalue $\jmath_{\scriptsize min}(\hat{\boldsymbol\lambda})$
directly, either analytically or numerically, since this implies the repeated calculation of the lowest eigenvalue of the large matrix
${\mathbbm J}({\boldsymbol\lambda})$.
In this situation, one can hope to split ${\mathbbm J}({\boldsymbol\lambda})$ into smaller blocks by exploiting its symmetry.
In fact, ${\mathbbm J}({\boldsymbol\lambda})$ commutes with the lattice translation operators $T_{\mathbf a},\, {\mathbf a}\in{\mathbbm Z}_{\mathbf N}$
and hence both operators possess a common system of eigenvectors.
The eigenvectors of $T_{\mathbf a}$ form the discrete Fourier basis $\exp ({\sf i}\,{\mathbf n}\cdot{\mathbf q})$.
Hence we seek for eigenvectors $t$ of ${\mathbbm J}(\hat{\boldsymbol\lambda})$ of the product form
\begin{equation}\label{eigFourier}
  t_{i{\mathbf n}}= z_i\, {\sf e}^{{\sf i}\,{\mathbf n}\cdot{\mathbf q}}
  \;,
\end{equation}
 where the $z_i$ are the components of a vector ${\mathbf z}\in{\mathbbm C}^L$ and
 the ``wave vector"
 ${\mathbf q}$, runs through the
 finite ``Brillouin set"  ${\mathcal Q}$ defined by
 \begin{eqnarray}\nonumber
  {\mathcal Q}&:=& \Big\{\left( q_1,\ldots, q_d\right) \left| \right.\Big.\\
 \nonumber
  &&\Big.q_i = \frac{2\pi k_i}{N_i}\mbox{ and } k_i\in{\mathbbm Z}_{N_i} \mbox{ for all }i=1,\ldots,d \Big\}
  ,\\
  \label{defQ} &&
 \end{eqnarray}
forming coordinates for certain points of the first Brillouin zone.
 Inserting the ansatz (\ref{eigFourier}) into (\ref{eiglatt}) gives
 \begin{eqnarray}
\nonumber
  && \sum_{j{\mathbf m}}{\mathbbm J}_{{\mathbf n},{\mathbf m}}^{ij}(\hat{\boldsymbol\lambda})\, z_j\, {\sf e}^{{\sf i}\,{\mathbf m}\cdot{\mathbf q}}\\
  \label{eigFour2}
    &\stackrel{(\ref{latticetrans})}{=}&
  \sum_{j{\mathbf m}}{\mathbbm J}_{{\mathbf 0}, {\mathbf m}-\mathbf {n}}^{ij}(\hat{\boldsymbol\lambda})\, z_j\,
  {\sf e}^{{\sf i}\,({\mathbf m}-{\mathbf n})\cdot{\mathbf q}}\,{\sf e}^{{\sf i}\,{\mathbf n}\cdot{\mathbf q}}\\
  \label{eigFour3}
  &=&
 \sum_j \underbrace{\left(\sum_{{\boldsymbol \ell}}{\mathbbm J}_{{\mathbf 0}, {\boldsymbol \ell}}^{ij}(\hat{\boldsymbol\lambda})\,
  {\sf e}^{{\sf i}\,{\boldsymbol \ell}\cdot{\mathbf q}}\right)}
  \,z_j\,{\sf e}^{{\sf i}\,{\mathbf n}\cdot{\mathbf q}}\\
   \label{eigFour4}
  &=& \sum_j \quad\quad \widehat{\mathbbm J}^{ij}(\hat{\boldsymbol\lambda},{\mathbf q})\quad\quad\quad z_j\,{\sf e}^{{\sf i}\,{\mathbf n}\cdot{\mathbf q}}
  \;,
 \end{eqnarray}
introducing the discrete Fourier-transformed $J$-matrix
\begin{equation}\label{FourJ}
  \widehat{\mathbbm J}^{ij}({\boldsymbol\lambda},{\mathbf q}):=
  \sum_{{\boldsymbol \ell}}{\mathbbm J}_{{\mathbf 0}, {\boldsymbol \ell}}^{ij}({\boldsymbol\lambda})\,
  {\sf e}^{{\sf i}\,{\boldsymbol \ell}\cdot{\mathbf q}}
  \;.
\end{equation}
Hence (\ref{eigFourier}) will be an eigenvector of ${\mathbbm J}(\hat{\boldsymbol\lambda})$ if
${\mathbf z}$ is chosen as an eigenvector of $\widehat{\mathbbm J}(\hat{\boldsymbol\lambda},{\mathbf q})$, i.~e.,
\begin{equation}\label{eigJF}
 \sum_j \widehat{\mathbbm J}^{ij}(\hat{\boldsymbol\lambda},{\mathbf q})\; z_j = \jmath(\hat{\boldsymbol\lambda},{\mathbf q})\,z_i
 \;,
\end{equation}
for all $i=1,\ldots,L$. In this context the following Proposition will be of interest:
\begin{prop}\label{P1}
Under the preceding conditions the following holds
for all ${\boldsymbol\lambda}\in\Lambda$ and ${\mathbf q}\in {\mathcal Q}$:\\
\begin{enumerate}
  \item $\widehat{\mathbbm J}({\boldsymbol\lambda},{\mathbf q})$ is an Hermitean  $L\times L$-matrix.
\item $\widehat{\mathbbm J}({\boldsymbol\lambda},{\mathbf q})=\widehat{\mathbbm J}^\top({\boldsymbol\lambda},-{\mathbf q})$,
where $^\top$ denotes the transposition of a matrix.
  \item $\widehat{\mathbbm J}({\boldsymbol\lambda},{\mathbf q})$ and $\widehat{\mathbbm J}({\boldsymbol\lambda},-{\mathbf q})$
  have the same eigenvalues and complex-conjugate eigenvectors.
\end{enumerate}
\end{prop}

The proof of this Proposition can be found in the Appendix \ref{sec:A1}.

 The advantage of considering the Fourier transform $ \widehat{\mathbbm J}({\boldsymbol\lambda},{\mathbf q})$
 for the calculation of the maximum of ${\boldsymbol\lambda}\mapsto \jmath_{\scriptsize min}({\boldsymbol\lambda})$
 is that we do not need to diagonalize an $M\times M$-matrix but only an $L\times L$-matrix, albeit for different values of
 ${\mathbf q}$.
 Since typically the number of spin sites $L$ in the unit cell is small, the diagonalization of the matrix
 $ \widehat{\mathbbm J}({\boldsymbol\lambda},{\mathbf q})$ is straightforward, sometimes it is even analytically possible.
 Let us reformulate the condition for the ground state gauge  $\hat{\boldsymbol\lambda}$ in this setting.
 Denote by $\jmath_{\scriptsize min}({\boldsymbol\lambda},{\mathbf q})$ the minimal eigenvalue of
 $\widehat{\mathbbm J}({\boldsymbol\lambda},{\mathbf q})$, then the ground state gauge  $\hat{\boldsymbol\lambda}$ will be
 uniquely determined by the condition
(``Max-Min-Principle")
 \begin{equation}\label{groundgauge}
  \jmath_{\scriptsize min}(\hat{\boldsymbol\lambda})=\text{Max}_{{\boldsymbol\lambda}\in{\Lambda}} \text{Min}_{{\mathbf q}\in{\mathcal Q}} \;
  \jmath_{\scriptsize min}({\boldsymbol\lambda},{\mathbf q})
  \;.
 \end{equation}

 In the special case of a Bravais lattice, i.~e., $L=|{\mathcal L}|=1$, the matrix $\widehat{\mathbbm J}({\boldsymbol\lambda},{\mathbf q})$
 reduces to a real number and necessarily ${\boldsymbol\lambda}={\mathbf 0}$ due to (\ref{sumlambda}).
 The minimal energy (per site) is obtained by the minimum of
 $\widehat{\mathbbm J}({\mathbf 0},{\mathbf q})$ over ${\mathbf q}\in {\mathcal Q}$. The corresponding ground state is the two-dimensional spiral
 state given by the real and imaginary part of the Fourier basis $\exp\left( {\sf i}{\mathbf n}\cdot{\mathbf q}\right)$.
 In this way, we recover the LTLK-solution of the ground state problem for Bravais spin lattices, see \cite{LK60,KM07}.

 Returning to general spin lattices we consider the case of three-dimensional ground states in some detail, the other two cases
 being analogous but simpler. We thus assume that we have found three linearly independent eigenvectors ${\mathbf z}^{(k)},\;k=1,2,3,$
 of $\widehat{\mathbbm J}({\boldsymbol\lambda},{\mathbf q}^{(k)})$, resp.~, that will be used to form the three columns of a real matrix $W$,
 hence satisfying either
\begin{equation}\label{defWr}
 W_{i,k}={\mathbf z}^{(k)}_i,\; \mbox{for }  i=1,\ldots,L
 \;.
 \end{equation}
 if ${\mathbf z}^{(k)}$ is real, or
\begin{eqnarray}\label{defWc1}
 W_{i,k}
  &=&
 \Re\left({\mathbf z}^{(k)}_i\right),\; \mbox{for }  i=1,\ldots,L
 \\
 \label{defWc2}
 W_{i,k+1}
  &=&
 \Im\left({\mathbf z}^{(k)}_i\right),\; \mbox{for }  i=1,\ldots,L
 \;,
\end{eqnarray}
 if ${\mathbf z}^{(k)}$ is complex. In the latter case we use the convention ${\mathbf q}^{(k+1)}=-{\mathbf q}^{(k)}$
and rely on Proposition \ref{P1}{.3}.

We will consider linear combinations of these eigenvectors that give rise to ground state spin configurations in the primitive unit cell of the form
${\mathbf s}=W\,\Gamma$, where $\Gamma$ is a real $3\times 3$-matrix representing the coefficients of the linear combinations.
The linear combination represented by $\Gamma$ will be called {\em admissible} iff $\Gamma_{k,\ell}=0$
holds in case of ${\mathbf q}^{(k)}\neq {\mathbf q}^{(\ell)}$. In other words: Admissible linear combinations
do not mix eigenvectors with different ${\mathbf q}$-vectors and also do not mix the real and imaginary part of a complex eigenvector.

Then we have the following result:
\begin{prop}\label{P2}
 If there exists an admissible linear combination $\Gamma$ that yields a ground state configuration
 $\left( {\mathbf s}_{i,{\mathbf 0}}\right)_{i\in{\mathcal L}}$
  of three-dimensional unit vectors in the primitive unit cell
 then it can be extended to a total ground state spin configuration
 $\left( {\mathbf s}_{i,{\mathbf n}}\right)_{i\in{\mathcal L},{\mathbf n}\in{\mathbbm Z}_{\mathbf N}}$
 that will also consist of unit vectors.
\end{prop}
The proof of this Proposition and the detailed form of the extension can be found in Appendix \ref{sec:A2}.

We will illustrate our approach for the above example of cyclic sawtooth chain simplified to $J_1=J_2=J_3=1$,
see Figure \ref{FIGCP},
{left panel,}
although this system is already
 rather small  with $M=12$. The  Fourier transform $\widehat{\mathbbm J}(\lambda, q)$ assumes the form
 \begin{equation}\label{FC6}
  \widehat{\mathbbm J}(\lambda,q)=
  \left(
\begin{array}{cc}
 \lambda +{\sf e}^{-{\sf i} q}+{\sf e}^{{\sf i} q} & 1+{\sf e}^{-{\sf i} q} \\
 1+{\sf e}^{{\sf i} q} & -\lambda  \\
\end{array}
\right)
\;.
\end{equation}
We mention that we may choose a large value for $M$, the total number of spins,
which would not change the matrix (\ref{FC6}) but only increase the
number of $q$-values to be considered.
We have plotted the four functions $\lambda\mapsto \jmath_{\scriptsize min}(\lambda,q)$ corresponding
to $q=0,\pm\frac{\pi}{3},\pm\frac{2\pi}{3},\pi$, see Figure \ref{FIGCP}, { right panel,} where the two signs of, say,
$\pm\frac{\pi}{3}$ yield the same function $\jmath_{\scriptsize min}(\lambda,q)$ according to Proposition \ref{P1}.
We see that $\jmath_{\scriptsize min}(\lambda)$ which is the minimum of these four functions has a unique
maximum at $\hat{\lambda}=1/2$ of the height $\jmath_{\scriptsize min}(1/2)=-3/2$, in accordance with
(\ref{groundgauge}).
This corresponds to the ground state energy of $E_{\scriptsize min}=M \,\jmath_{\scriptsize min}(1/2)=-18$
that can be realized by any spin configuration with an angle of $120^\circ$ between adjacent spins, see below.
Moreover, we observe that the function $\lambda\mapsto \jmath_{\scriptsize min}(\lambda,\pm\frac{2\pi}{3})$
(green curve in Figure \ref{FIGCP}, { right panel}) has a smooth maximum at the critical point.
This gives rise to a special two-dimensional ground state as we will explain for the general case in the following subsection.

\subsection{Smooth maximum case and spiral ground states}\label{sec:SM}

In this subsection we assume that for some fixed ${\mathbf q}\in{\mathcal Q}$ the eigenvalue
$\jmath_{\scriptsize min}({\boldsymbol \lambda},{\mathbf q})$ is non-degenerate in some neighbourhood
of the critical point at ${\boldsymbol \lambda}=\hat{\boldsymbol \lambda}$ satisfying
$\jmath_{\scriptsize min}(\hat{\boldsymbol \lambda})=\jmath_{\scriptsize min}(\hat{\boldsymbol \lambda},{\mathbf q})$
and that the function
${\boldsymbol \lambda}\mapsto \jmath_{\scriptsize min}({\boldsymbol \lambda},{\mathbf q})$
has a smooth maximum at the critical point. This entails
\begin{equation}\label{partialder}
 \left.\frac{\partial}{\partial \lambda_i}\jmath_{\scriptsize min}({\boldsymbol \lambda},{\mathbf q})\right|_{{\boldsymbol \lambda}=\hat{\boldsymbol \lambda}}
 =0
\end{equation}
for $i=1,\ldots,L-1$. The latter restriction is due to the constraint $\sum_{i=1}^{L}\lambda_i =0$ and hence only the first
$L-1$ components of ${\boldsymbol\lambda}$ can be chosen independently, but $\lambda_L=-\sum_{i=1}^{L-1}\lambda_i$.
We consider a vector ${\mathbf z}\in {\mathbbm C}^L$ satisfying the eigenvalue equation (\ref{eigJF})
and being normalized according to $\sum_i \overline{z_i}\,z_i=1$. Rewrite (\ref{eigJF})  as
\begin{equation}\label{eig1}
  \sum_j \widehat{\mathbbm J}^{ij}({\mathbf 0},{\mathbf q})\,z_j =\left(\jmath_{\scriptsize min}({\boldsymbol \lambda},{\mathbf q})-\lambda_i \right)z_i
  \;,
\end{equation}
which implies
\begin{equation}\label{eig2}
  \sum_{ij} \overline{z_i}\,\widehat{\mathbbm J}^{ij}({\mathbf 0},{\mathbf q})\,z_j
  =\jmath_{\scriptsize min}({\boldsymbol \lambda},{\mathbf q})
  \underbrace{\sum_i\overline{z_i}\,z_i}_{=1}-\sum_{i=1}^{L}\lambda_i\,\overline{z_i}\,z_i
  \;.
\end{equation}

\begin{figure}[htp]
\centering
\includegraphics[width=1.0\linewidth]{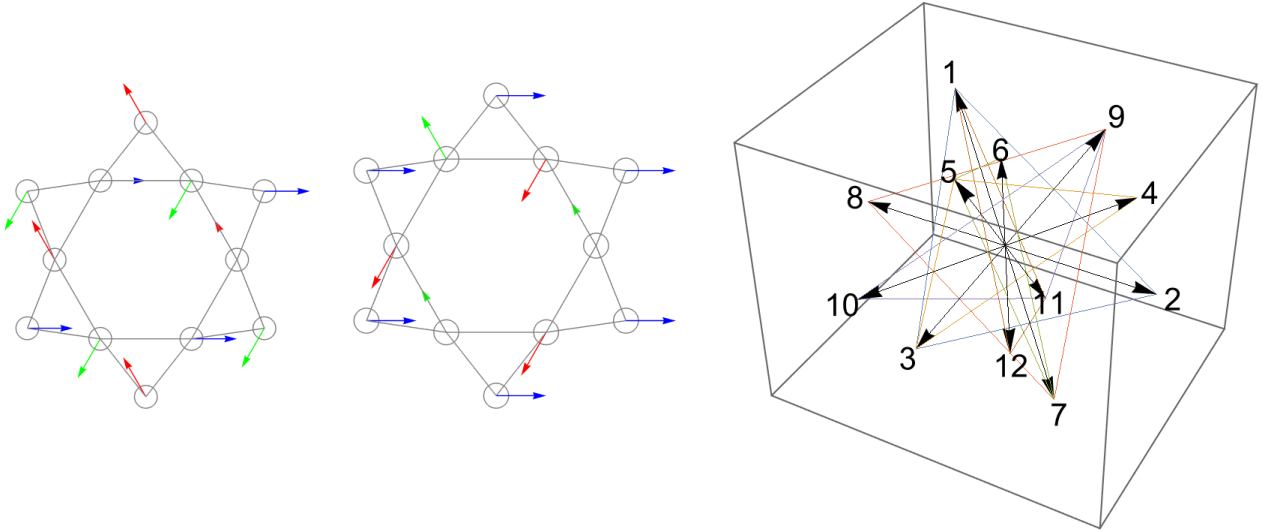}
\caption{Left panel: Plot of a spiral ground state of the cyclic sawtooth chain with $J_1=J_2=J_3=1$ according to (\ref{s12}).\\
Middle panel: Plot of another two-dimensional ground state of the cyclic sawtooth chain with $J_1=J_2=J_3=1$ according to (\ref{s3}).
\\
Right panel: The $3$-dimensional ground state of the cyclic sawtooth chain according to (\ref{ground3}).
The numbers refer to Figure \ref{FIGCP}, {left panel}.
The spin vectors ${\mathbf s}_\mu$ for $\mu=1,2,3,7,8,9$ lie in a certain common plane
$P_1$, analogous for $\mu=3,4,5,9,10,11$ ($P_2$) and $\mu=5,6,7,11,12,1$ ($P_3$).
}
\label{FIGCG}
\end{figure}

Since the l.~h.~s.~of (\ref{eig2}) is independent of ${\boldsymbol\lambda}$ the partial derivatives (\ref{partialder}) yield
\begin{equation}\label{eig3}
 0\stackrel{(\ref{partialder})}{=}\left.\frac{\partial}{\partial \lambda_i}\jmath_{\scriptsize min}({\boldsymbol \lambda},{\mathbf q})\right|_{{\boldsymbol \lambda}=\hat{\boldsymbol \lambda}} = \overline{z_i}\,z_i- \overline{z_L}\,z_L
 \;,
\end{equation}
for $i=1,\ldots,L-1$.
This means that the eigenvector ${\mathbf z}\in {\mathbbm C}^L$ of $\widehat{\mathbbm J}(\hat{\boldsymbol\lambda},{\mathbf q})$
has components of constant absolute values. Let us write $z_i=x_i+{\sf i}y_i$ for $i=1,\ldots,L$ and rescale  ${\mathbf z}$ such that
\begin{equation}\label{rescale}
 \left| z_i\right|^2 = x_i^2 + y_i^2 =1 \mbox{ for all } i=1,\ldots,L
 \;.
\end{equation}
We thus obtain two linearly independent eigenvectors of ${\mathbbm J}(\hat{\boldsymbol\lambda},{\mathbf q})$,
resp.~${\mathbbm J}(\hat{\boldsymbol\lambda},{-\mathbf q})$, of the form
\begin{equation}\label{eigen1}
  t_{i{\mathbf n}} = (x_i+{\sf i}\,y_i)\, {\sf e}^{{\sf i}\,{\mathbf n}\cdot{\mathbf q}}
  \;,
\end{equation}
and, using Proposition \ref{P1},
\begin{equation}\label{eigen2}
 \overline{ t_{i{\mathbf n}}} = (x_i-{\sf i}\,y_i)\, {\sf e}^{-{\sf i}\,{\mathbf n}\cdot{\mathbf q}}
  \;.
\end{equation}
From these we can form two linearly independent {\em real} eigenvectors of the form
\begin{equation}\label{real1}
 {\mathbf s}_{i{\mathbf n}}^{(1)}=\frac{1}{2}\left(t_{i{\mathbf n}}+\overline{ t_{i{\mathbf n}}} \right)=
 x_i\,\cos({\mathbf n}\cdot{\mathbf q})-y_i\,\sin({\mathbf n}\cdot{\mathbf q})
 \;,
\end{equation}
and
\begin{equation}\label{real2}
 {\mathbf s}_{i{\mathbf n}}^{(2)}=\frac{1}{2{\sf i}}\left(t_{i{\mathbf n}}-\overline{ t_{i{\mathbf n}}} \right)=
 x_i\,\sin({\mathbf n}\cdot{\mathbf q})+y_i\,\cos({\mathbf n}\cdot{\mathbf q})
 \;.
\end{equation}
The two-dimensional spin configuration $ {\mathbf s}_{i{\mathbf n}}:=\left({\mathbf s}_{i{\mathbf n}}^{(1)} ,{\mathbf s}_{i{\mathbf n}}^{(2)}\right)$
consists of unit vectors according to
\begin{equation}\label{unitvectors}
 {\mathbf s}_{i{\mathbf n}}\cdot {\mathbf s}_{i{\mathbf n}} = \left( {\mathbf s}_{i{\mathbf n}}^{(1)}\right)^2+\left({\mathbf s}_{i{\mathbf n}}^{(2)}\right)^2
  =x_i^2+y_i^2 \stackrel{(\ref{rescale})}{=}1
  \;,
\end{equation}
for all $i=1,\ldots,L$ and ${\mathbf n}\in {\mathbbm Z}_{\mathbf N}$, and hence can be considered as a ``spiral" ground state of the spin lattice.\\

We will evaluate this construction of a ground state for the above example of a cyclic sawtooth chain. For $\hat{\lambda}=1/2$ and
$q=2\pi/3$ the Fourier transformed $J$-matrix (\ref{FC6}) assumes the form
\begin{equation}\label{FJ1}
 \widehat{\mathbbm J}\left( \frac{1}{2},\frac{2\pi}{3}\right) =
 \left(
\begin{array}{cc}
 -\frac{1}{2} & \frac{1}{2}-\frac{{\sf i} \sqrt{3}}{2} \\
 \frac{1}{2}+\frac{{\sf i} \sqrt{3}}{2} & -\frac{1}{2} \\
\end{array}
\right)
\;.
\end{equation}
Its normalized eigenvector corresponding to the eigenvalue $\jmath_{\scriptsize min}\left( \frac{1}{2},\frac{2\pi}{3}\right)=-\frac{3}{2}$
will be
\begin{equation}\label{eigjmin}
 \left( x_1+{\sf i}y_1,x_2+{\sf i}y_2\right)=\left(-\frac{1}{2 \sqrt{2}}+\frac{\sf i}{2} \sqrt{\frac{3}{2}},\frac{1}{\sqrt{2}}\right)
 \;.
\end{equation}
The two eigenvectors of  ${\mathbbm J}(\hat{\lambda},{q})$ corresponding to (\ref{real1}) and (\ref{real2})
form the columns of the spin configuration matrix

\begin{equation}\label{s12}
{\mathbf s}=
\left(
\begin{array}{cccccccccccc}
  -\frac{1}{2} &1 &  -\frac{1}{2} &   -\frac{1}{2} &  1 &   -\frac{1}{2} &   -\frac{1}{2} &  1 &    -\frac{1}{2} &   -\frac{1}{2} &  1 &  -\frac{1}{2}\\
  \frac{\sqrt{3}}{2}& 0 &-\frac{\sqrt{3}}{2} & \frac{\sqrt{3}}{2} & 0& -\frac{\sqrt{3}}{2}& \frac{\sqrt{3}}{2}& 0& 0 & -\frac{\sqrt{3}}{2}& 0& -\frac{\sqrt{3}}{2}
 \end{array}
\right)^{{\top}}
\end{equation}
and give rise to the spiral ground state with angles of $120^\circ$ between adjacent spins depicted in Figure \ref{FIGCG},
{left panel}.
This ground state can of course be found directly in a simple way;
we just wanted to demonstrate that it also arises as a result of the theory presented here.

There is a second two-dimensional ground state where the spins in the inner hexagon point alternately in two different directions,
which are represented by the colors red and green in Figure \ref{FIGCG},
{middle panel,}
while the outer spins constantly point in the third direction,
which is represented by the color blue.
It remains a task to obtain the latter from the present theory.

\subsection{Other two-dimensional ground states}\label{sec:OC}

The spiral ground state configurations considered in the last subsection have been obtained as superpositions of a complex eigenvector
$z_i\,{\sf e}^{{\sf i}{\mathbf q}\cdot{\mathbf n}}$ of ${\mathbbm J}(\hat{\boldsymbol\lambda})$
and its complex conjugate $\overline{z}_i\,{\sf e}^{-{\sf i}{\mathbf q}\cdot{\mathbf n}}$. Another possibility to construct
real two-dimensional ground states would be a suitable superposition of two {\em real} eigenvectors $t^{(1)}$ and $t^{(2)}$.
Real eigenvectors of ${\mathbbm J}(\hat{\boldsymbol\lambda})$ occur for ${\sf e}^{{\sf i}\,{\mathbf n}\cdot{\mathbf q}}$ being real,
that is, for wave vectors ${\mathbf q}\in{\mathcal Q}$ having only components of the form $0$ or $\pm \pi$.
Additional conditions are that the corresponding eigenvalue must be $\jmath_{\scriptsize min}(\hat{\boldsymbol\lambda})$
and that $\jmath_{\scriptsize min}(\hat{\boldsymbol\lambda})$ is the maximum of all perturbed eigenvalues $\jmath({\boldsymbol\lambda})$
for ${\boldsymbol\lambda}$ in some neighbourhood of $\hat{\boldsymbol\lambda}$. Since the case of a smooth maximum has already be
treated in subsection \ref{sec:SM} we are left with the occurrence of a singular maximum in the form of a double cone or a wedge, see \cite{S17a}.

Let us denote by $W$ the real $M\times 2$-matrix with the two columns formed by the two eigenvectors $t^{(1)}$ and $t^{(2)}$.
The two components of the ground state configuration ${\mathbf s}_\mu,\,\mu\in{\mathcal M}$ are obtained as the superpositions
\begin{equation}\label{groundstatesuper}
  {\mathbf s}_\mu^{(i)} = \sum_{j=1}^{2} W_{\mu}^{j} \Gamma_{j i}
\end{equation}
for all $\mu\in{\mathcal M}$  and $i=1,2$, in matrix notation
\begin{equation}\label{WGamma}
  {\mathbf s}=W\,\Gamma
\end{equation}
with a real $2\times 2$-matrix $\Gamma$. The corresponding {\em Gram matrix} $G$ of all scalar products between spin vectors
is given by
\begin{eqnarray}\label{Gram1}
  G={\mathbf s}\,{\mathbf s}^\top & =& W\,\left(\Gamma\,\Gamma^\top\right)\,W^\top =: W\,\Delta\,W^\top \\
  \label{Gram2}
  &=:& W\,
  \left( \begin{array}{cc}
           \delta_1 & \delta_2 \\
           \delta_2 & \delta_3
         \end{array}
  \right)\,W^\top
  \;.
\end{eqnarray}
The matrix $\Delta$ must be positively semi-definite, i.~e., $\delta_1,\delta_3\ge 0$ and $\delta_1\,\delta_3\ge \delta_2^2$,
and its entries have to be chosen such that
\begin{equation}\label{ADE}
 1=G_{\mu\mu}= \left( W\,\Delta\,W^\top \right)_{\mu\mu}\quad \mbox{ for all } \mu\in{\mathcal M}
 \;.
\end{equation}
The above conditions guarantee that there always exist solutions satisfying  $\Delta\ge 0$ and (\ref{ADE}), see \cite{S17a}.
Using the polar decomposition the matrix $\Gamma$ of superposition coefficients can be written as
\begin{equation}\label{Gamma1}
 \Gamma=\sqrt{\Delta}\,R
 \;,
\end{equation}
with an arbitrary rotation/reflection matrix $R\in O(2)$.\\

We will illustrate the foregoing considerations by choosing the real eigenvectors  $t^{(1)}$ and $t^{(2)}$ of ${\mathbbm J}(\hat{\boldsymbol\lambda})$
corresponding to the wave numbers $q=0$ and $q=\pm \pi$ of the cyclic sawtooth system. These eigenvectors assume the form
\begin{eqnarray}
\label{eigent1}
  t^{(1)} &=& (-1,2,-1,2,-1,2,-1,2,-1,2,-1,2)\;, \\
  \label{eigent2}
  t^{(2)} &=&(1,0,-1,0,1,0,-1,0,1,0,-1,0)
  \;.
\end{eqnarray}
The corresponding equation (\ref{ADE}) has the solution
\begin{equation}\label{soldelta}
 \delta_1=\frac{1}{4},\quad \delta_2=0,\quad \delta_3=\frac{3}{4}
 \;,
\end{equation}
and ${\mathbf s}=W\,\sqrt{\Delta}$ has the form
\begin{equation}\label{s3}
{\mathbf s}=\left(
\begin{array}{cccccccccccc}
  -\frac{1}{2} &  1 &  -\frac{1}{2} & 1 &  -\frac{1}{2} & 1 &  -\frac{1}{2} &  1 & -\frac{1}{2} & 1 & -\frac{1}{2} &  1 \\
   \frac{\sqrt{3}}{2}&0 &  -\frac{\sqrt{3}}{2} & 0 & \frac{\sqrt{3}}{2}  & 0 & - \frac{\sqrt{3}}{2} & 0 & \frac{\sqrt{3}}{2}& 0&  -\frac{\sqrt{3}}{2} & 0
\end{array}
\right)^{{\top}}
\;,
\end{equation}
see Figure \ref{FIGCG}, {middle panel}.

\subsection{Three-dimensional ground states}\label{sec:TG}

The procedure to obtain $3$-dimensional ground states is analogous to that sketched in subsection \ref{sec:OC}.
We consider a real $3$-dimensional subspace of the eigenspace of ${\mathbbm J}(\hat{\boldsymbol\lambda})$ corresponding
to the eigenvalue $\jmath_{\scriptsize min}(\hat{\boldsymbol\lambda})$. The components of the ground state are obtained as
linear combinations of these eigenvectors.
They either correspond to wave vectors ${\mathbf q}\in{\mathcal Q}$ having only components of the form $0$ or $\pm \pi$
or can be superposed from complex eigenvectors corresponding to ${\mathbf q}\in{\mathcal Q}$ and $-{\mathbf q}$.
$W$ is the $M\times 3$-matrix the columns of which are the chosen three real eigenvectors and $\Delta$ is a positively semi-definite
$3\times 3$-matrix satisfying the equation analogous to (\ref{ADE}).
In general, the solutions for $\Delta$ will form a $K$-dimensional convex set, see \cite{S17a}, and the matrix $\Gamma$ of superposition coefficients
will be given by (\ref{Gamma1}), where $R\in O(3)$.

To illustrate this construction we consider the $3$-dimensional subspace spanned by the eigenvectors of ${\mathbbm J}(\hat{\boldsymbol\lambda})$
corresponding to the wave numbers $q=\pi/3,\,q=- \pi/3$ and $q=\pi$ of the cyclic sawtooth chain
and the corresponding real subspace spanned by
\begin{eqnarray}
\label{eigt1}
  t^{(1r)} &=&(-1,2,-1,1,0,-1,1,-2,1,-1,0,1)\;,\\
\label{eigt2}
  t^{(1i)} &=& (1,0,-1,3,-2,3,-1,0,1,-3,2,-3)\;, \\
  \label{eigt3}
  t^{(2)} &=&(1,0,-1,0,1,0,-1,0,1,0,-1,0)
  \;.
\end{eqnarray}
Again, $W$ is the $12\times 3$-matrix formed of the three columns (\ref{eigt1}-\ref{eigt3}).
The corresponding equation (\ref{ADE}) has the unique solution
\begin{equation}\label{solDelta}
 \Delta=\left(
\begin{array}{ccc}
 \frac{1}{4} & 0 & 0 \\
 0 & \frac{1}{12} & 0 \\
 0 & 0 & \frac{2}{3} \\
\end{array}
\right)
\;.
\end{equation}
Then we obtain the $3$-dimensional ground state
\begin{equation}\label{ground3}
{\mathbf s}=W\,\sqrt{\Delta}=
 \left(
\begin{array}{cccccccccccc}
 -\frac{1}{2} & 1 & -\frac{1}{2} & \frac{1}{2} & 0 &  -\frac{1}{2} & \frac{1}{2} & -1 & \frac{1}{2} & -\frac{1}{2} & 0 & \frac{1}{2} \\
  \frac{1}{2 \sqrt{3}} &0 &  -\frac{1}{2 \sqrt{3}} &  \frac{\sqrt{3}}{2} &-\frac{1}{\sqrt{3}} &\frac{\sqrt{3}}{2} &  -\frac{1}{2 \sqrt{3}} & 0 &
   \frac{1}{2 \sqrt{3}} & -\frac{\sqrt{3}}{2} & \frac{1}{\sqrt{3}} & -\frac{\sqrt{3}}{2}\\
  \sqrt{\frac{2}{3}} &  0 & -\sqrt{\frac{2}{3}} &0 &  \sqrt{\frac{2}{3}} &  0 & -\sqrt{\frac{2}{3}}& 0 & \sqrt{\frac{2}{3}} & 0& -\sqrt{\frac{2}{3}}& 0
 \end{array}
\right)^{{\top}}
\;,
\end{equation}
shown in Figure \ref{FIGCG}, {right panel}.
For this state the six spin triangles have local ground states
lying in three different planes $P_1,\, P_2,\,P_3$ that are related by rotations  with  the angle of $120^\circ$ about
a constant axis, here chosen as the $3$-axis.

\subsection{Infinite lattices and incommensurable ground states}\label{sec:IL}
Our theory as outlined in the preceding Sections \ref{sec:GS} - \ref{sec:TG} is, in principle, restricted to finite models of an infinite lattice.
However, it may happen that the ``true" ground state of an infinite lattice cannot be obtained by finite models but only approximated.
Such ground states have been called ``incommensurable" in the literature, see, e.~g., \cite{N04,Ketal00}.
Although they are, strictly speaking, beyond the present theory, there is a chance to obtain incommensurable states by an
extrapolation of the Max-Min-Principle $(\ref{groundgauge})$ to
\begin{equation}\label{groundgaugeinfinite}
  \jmath_{\scriptsize min}(\hat{\boldsymbol\lambda})=\text{Max}_{{\boldsymbol\lambda}\in{\Lambda}} \text{Min}_{{\mathbf q}\in{\mathcal Q}^\infty} \;
  \jmath_{\scriptsize min}({\boldsymbol\lambda},{\mathbf q})
  \;,
 \end{equation}
 where the finite Brillouin set
 ${\mathcal Q}$ has been extended to the infinite one:
 \begin{equation}\label{defQinf}
   {\mathcal Q}^\infty:=\left[-\pi,\pi \right]^d
   \;.
 \end{equation}
 An example will be given in Section \ref{sec:MSL}.

\section{Applications}\label{sec:AP}

\subsection{General remarks}\label{sec:GR}

Having presented a generalization of the LTLK approach, it is natural to look for applications
where our approach provides ground states that cannot be obtained with LTLK theory,
beyond the toy example of the cyclic sawtooth chain.
In doing so, we must take into account the fact that the LTLK theory can fail for various reasons.
One reason is the presence of non-equivalent spins in a non-Bravais lattice as in the cyclic sawtooth chain.
Another possible problem could be that LTLK theory provides all the mathematical ground states,
but {they might} be unphysical because their dimension is larger than three.
{We have already mentioned this problem at the end of Section \ref{sec:GS}.
We would like to add here only the remark that ground states with minimal dimension $>3$
can occur also in spin lattices, even quite often, and are a problem for our theory as well as for LTLK theory.
}

An example is the $J_1 - J_3-$ kagome lattice, where the Luttinger-Tizsa lower bound is reached everywhere except
in a small ''gray" region, see Fig.~$2$ in \cite{Getal20}. We have studied this example and found
{that the ground state which assumes the mentioned lower bound is  $6$-dimensional.}
However, { as we have already mentioned,}
our approach is not tailored to find physical, i.~e., at most $3$-dimensional,
ground states when the mathematical ground states are $K$-dimensional with $K>3$, as in this example.
A similar situation occurs for the  $J_1 - J_3-$ pyrochlore lattice
which can be considered as a $3$-dimensional analogue of the kagome lattice. A numerical study of the pyrochlore ground states
with nearest and next-nearest neighbor coupling has been given in \cite{LH12}.

There is a second class of applications with non-equivalent spins in the primitive unit cell of a non-Bravais lattice,
where our approach yields all symmetric ground states, but these could also be obtained by elementary reasoning.
As an example we mention
a square-kagome lattice that has two non-equivalent sites as well as {two} different NN bonds $J_1, J_2$ \cite{Retal09}.
It is a system of corner-sharing triangles.
Each triangle has its known two-dimensional ground state, and these ground states can be composed
into the global ground state with the local states more or less free to rotate, see \cite{Retal09}.
Therefore, examples of this type would not illustrate the power of our method.

{We finally found two examples where the mentioned drawbacks do not occur:
First, the $J_1 -J_2-$honeycomb lattice, Section \ref{sec:AH}.
While there are one-dimensional ground states of elementary type here as well,
the construction of the remaining $3$-dimensional ground states can be seen as a successful application of our theory.
Second, in Sec. \ref{sec:MSL} we consider a modified square lattice with different ground state phases, including incommensurable ones.
Both examples have not been considered before in the literature and are deliberately chosen for our purposes.}

\subsection{Modified honeycomb lattice}\label{sec:AH}

\begin{figure}[htp]
\centering
\includegraphics[width=1.0\linewidth]{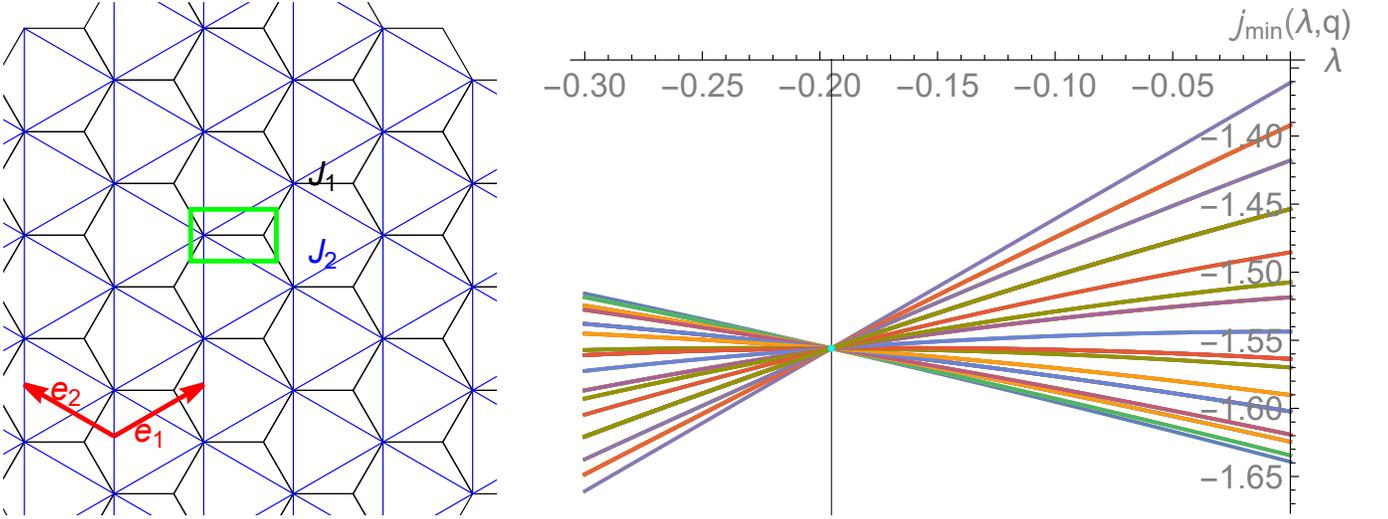}
\caption{
{
Left panel: Sketch of the $J_1-J_2$ honeycomb lattice. It is a non-Bravais lattice
that can be generated by multiples of translations of two spin sites in the primitive unit cell (green rectangle)
with translations vectors ${\mathbf e}_1$ and ${\mathbf e}_2$ (red vectors).
Due to the specific choice of the $J_2$ bonds
the two sites in the primitive unit cell become nonequivalent.}\\
{
Right panel: Plot of $\jmath_{\scriptsize min}(\lambda,{\mathbf q})$ for the modified honeycomb lattice,  and $85$ different values of ${\mathbf q}\in {\mathcal Q}$.
The angle $\phi$ is chosen as  $\phi=1.1$ such that the ground states belong to phase $II^\ast$, see Figure \ref{FIGSHE}, upper panel.
All curves cross at the critical point with coordinates
$(\hat{\lambda}=-0.19511,\jmath_{\scriptsize min}(\hat{\lambda})= -1.5559)$.
}
}
\label{FIGHPC}
\end{figure}

We consider the $J_1- J_2 -$honeycomb lattice that can be generated by two spin sites with coordinates, say, $(0,0)$ and $(1,0)$,
in the primitive unit cell
and integer multiples of translations into the directions ${\mathbf e}_1=\left(\frac{3}{2},\frac{\sqrt{3}}{2}\right)$
and  ${\mathbf e}_2=\left(\frac{3}{2},-\frac{\sqrt{3}}{2}\right)$, see Figure \ref{FIGHPC}, {left panel}.
We will denote the translates of $(0,0)$ as ``even sites" and those of  $(1,0)$ as ``odd sites".
Since the second neighbor bond $J_2$ connects only even sites the spins in {the} unit cell become non-equivalent.
The even sites form a triangular lattice with coupling constant $J_2$ between adjacent sites. Every second
triangle of this lattice is occupied by an odd spin that is coupled with strength $J_1$ to its three adjacent even spins,
see Figure \ref{FIGHPC}, {left panel}.
Since the ground states do not depend on a common positive factor of $J_1$ and $J_2$ we may set
\begin{equation}\label{setJ1J2}
J_1=\sin \phi\quad \mbox{ and } J_2=\cos \phi,\quad -\pi<\phi\le \pi
\;,
\end{equation}
without loss of generality.

From this it already follows that the simultaneous replacements $J_1\mapsto -J_1$
and inversion of all even spins leave the total energy (per site) invariant.
Hence the energy of the ground states $E_{\scriptsize min}(\phi)$ will be an even function of $\phi$.

The following calculations refer to a finite model of the $d=2$-dimensional honeycomb lattice of the kind (\ref{structureM}) with
a unit cell  ${\mathcal L}=\{0,1\}$ containing
$L=|{\mathcal L}|=2$ sites and $N_1\times N_2=6\times 6=36$
copies of the unit cell. At first sight, the finite Brioullin set ${\mathcal Q}$ would contain $N_1\times N_2 =36$ elements,
but due to $ \jmath_{\scriptsize min}({\mathbf q})= \jmath_{\scriptsize min}(-{\mathbf q})$ this set can be reduced to $20$ elements.
The $2\times 2$-matrix $\widehat{\mathbbm J}({\mathbf q},\lambda)$ is readily calculated as
\begin{equation}\label{Jq}
 \widehat{\mathbbm J}({\mathbf q},\lambda)=
 \left(
\begin{array}{cc}
J_{11} & J_{12}   \\
J_{21}& J_{22}  \\
\end{array}
\right)
\;,
\end{equation}
where
\begin{eqnarray}
\label{J11}
  J_{11} &=&2 \cos \phi  (\cos (q_1-q_2)+\cos q_1+\cos q_2) +\lambda \\
  \label{J12}
  J_{12} &=& \overline{J_{21}}=\sin \phi \left({\sf e}^{-{\sf i} q_1}+{\sf e}^{-{\sf i} q_2}+1\right)  \\
  \label{J22}
  J_{22}&=&-\lambda
  \;.
\end{eqnarray}
We see that $\widehat{\mathbbm J}({\mathbf q},\lambda)$ is invariant under the reflection $q_1 \leftrightarrow q_2$.

\begin{figure}[htp]
\centering
\includegraphics[width=1.0\linewidth]{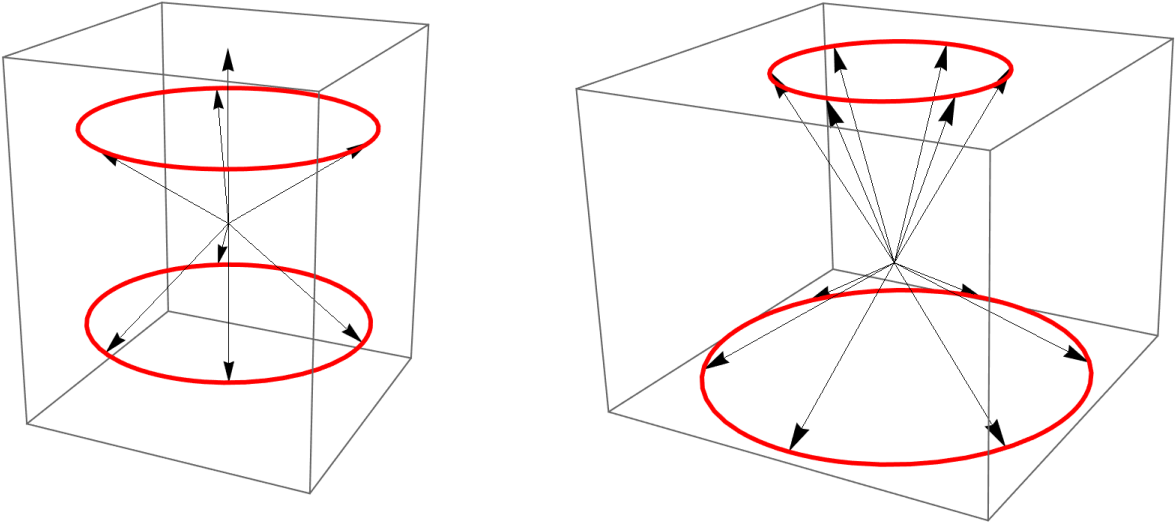}
\caption{
{
Left panel: Sketch of the $8$ spin vectors forming the ground state (\ref{groundstate1a}-\ref{groundstate1d}) of the $J_1 - J_2 -$honeycomb lattice
for the choice $\phi=\pi/6$. The two vectors $(0,0,\pm 1)$ are assumed by odd spin sites;
all other spin vectors are assumed by even sites and lie on two (red) circles with $3$-components $\pm\frac{1}{\sqrt{3}}$.
}\\
{
Right panel:
Sketch of the $12$ spin vectors forming the ground state (\ref{groundstate2c}-\ref{groundstate2f}) of the $J_1 - J_2 -$honeycomb lattice
for the choice $\phi=\pi/3$. They lie on two (red) circles with $3$-components $\frac{\sqrt{3}}{2}$ (odd sites) and $-\frac{1}{2}$ (even sites).
}
}
\label{FIGABC}
\end{figure}

\begin{figure}[htp]
\centering
\includegraphics[width=0.7\linewidth]{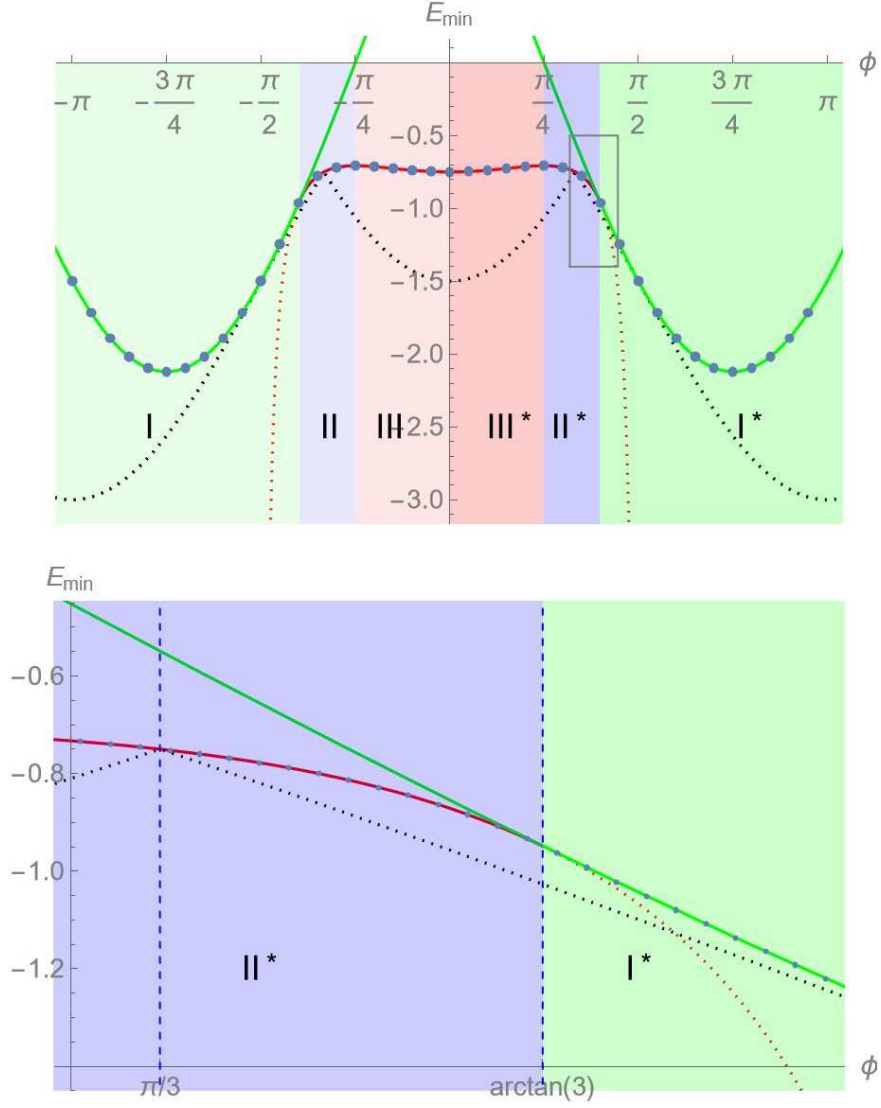}
\caption{Plot of a minimal energy of the $J_1-J_2$ honeycomb lattice as a function of $\phi$. The black dotted curve represents
the minimal eigenvalues of the $J$-matrix with $\lambda=0$ (lower Luttinger-Tisza bound).
The green curves correspond to $ E^{(\pm)}_{\scriptsize min}$ according to (\ref{energypm}),
and the red curve to $ E^{(1)}_{\scriptsize min}$ according to  (\ref{energy1}). The blue points represent the
numerical Monte Carlo results for the minimal energy.
The upper panel represents the global view, and the lower one the magnification of the rectangle in the upper panel. Especially the lower panel shows
that at $\phi=\pi/3$ the minimal energy $E_{\scriptsize min}=-3/2$ coincides with the minimal eigenvalue of the $J$-matrix, and that at
$\phi=\arctan 3$ the two functions $ E^{(+)}_{\scriptsize min}$ and  $ E^{(1)}_{\scriptsize min}$ intersect with the same slope.
All minimal energies are multiplied with a global factor of $1/2$ according to the convention used in large parts of the literature.
We distinguish between six phases ($I\; - \;III^\ast$) indicated by light colors as explained in the text.
}
\label{FIGSHE}
\end{figure}

The ground states will be ferromagnetic at least for $J_1, J_2<0$, i.~e., for the interval $-\pi <\phi < -\pi/2$.
This constitutes the ferromagnetic phase $I$.
According to the above remarks there exists an analogous
one-dimensional ground state phase (phase $I^\ast$) for the interval $\pi/2<\phi < \pi$
that is obtained by inverting all even spins of the ferromagnetic ground state
and hence represents an anti-ferromagnetic N\'{e}el state.
The corresponding ground state energy will be
\begin{equation}\label{energypm}
 E^{(\pm)}_{\scriptsize min}=3\left( \cos \phi \pm  \sin \phi\right)
 \;,
\end{equation}
where the sign $\pm$ has to be chosen positive for the interval $-\pi <\phi < -\pi/2$ and negative for  $\pi/2<\phi < \pi$.
Both types of "one-dimensional phases" are shown in light green in different gradations in Figure \ref{FIGSHE}.

These ground states can also be obtained as follows. The lowest eigenvalue of  $\widehat{\mathbbm J}({\mathbf 0},{\lambda})$
has a global smooth maximum at $\lambda=\hat{\lambda}=-3\,\cos\phi$ with maximal value (\ref{energypm}). The corresponding eigenvectors
are $(\pm 1,1)$ and its components are of absolute value $1$ in accordance with the considerations in subsection \ref{sec:SM}.
These eigenvectors yield the one-dimensional ground states via (\ref{eigJF}).
Next, we will follow the steps of analytically obtaining the ground states
of two other phases that prevail for the remaining values of $\phi$.

First, we choose a small value of $\phi$, say, $\phi=0.1$ and numerically calculate the maximum w.~r.~t.~$\lambda$ of the minima
of the eigenvalues of $J({\mathbf q},\lambda),\;{\mathbf q} \in{\mathcal Q}$. This maximum has the value $E_{\scriptsize min}=-1.49751$
and is obtained for $\hat{\lambda}=1.4875$. Next we ask for which ${\mathbf q} \in{\mathcal Q}$ this maximum is obtained.
The somewhat surprising answer is: for all ${\mathbf q} \in{\mathcal Q}$.
At first sight, this seems unfavorable since it means that, in principle, we would have to construct the ground state
as a superposition of $|{\mathcal Q}|=20$ eigenvectors. Fortunately, it is possible to find two values of ${\mathbf q} \in{\mathcal Q}$
that already suffice to construct the ground states, namely ${\mathbf q}_a=(0,\pi)$ and ${\mathbf q}_b=(\frac{2\pi}{3},-\frac{2\pi}{3})$.

As an aside, we add that the above degeneracy of $E_{\scriptsize min}$ is not due to the choice of a finite ${\mathcal Q}$:
it holds in general that $E_{\scriptsize min}$ is independent of ${\mathbf q}$ and thus represents a ``flat band".
We have thus found a new facet of the interesting field of flat-band physics
that has emerged in the last few decades, see, e.~g., \cite{M91a,M91b,Retal04,KMH20,LAF18}.

Coming back to our problem we note that the condition that two suitable eigenvalues of $\widehat{\mathbbm J}({\mathbf q}_a,\lambda)$ and
$\widehat{\mathbbm J}({\mathbf q}_b,\lambda)$ coincide at the global maximum leads
to an analytically determination of $\hat{\lambda}$ as
\begin{equation}\label{lambdamax}
 \hat{\lambda}=2 \cos \phi-\frac{\sec \phi}{2}
 \;,
\end{equation}
and the corresponding energy
\begin{equation}\label{energy1}
 E^{(1)}_{\scriptsize min}=- \cos \phi-\frac{1}{2} \sec\phi
 \;.
\end{equation}
The eigenvector of $\widehat{\mathbbm J}\left({\mathbf q}_a,\hat{\lambda}\right)$ corresponding to the eigenvalues (\ref{energy1})
will be ${\mathbf a}=(a_0,a_1)=(-\tan \phi,1)$.  $\widehat{\mathbbm J}\left({\mathbf q}_b,\hat{\lambda}\right)$ happens to be already diagonal
and the eigenvector corresponding to the eigenvalue (\ref{energy1}) will be ${\mathbf b}=(b_0,b_1)=(1,0)$.
For simplicity, we will use complex multiples of
${\mathbf b}$ to represent the components of spin vectors in the $x-y-$plane. Then it is straightforward to write the ground state configuration
as the following superposition of ${\mathbf a}\,{\sf e}^{{\sf i}{\mathbf n}\cdot{\mathbf q}_a}$ and
${\mathbf b}\,{\sf e}^{{\sf i}{\mathbf n}\cdot{\mathbf q}_b}$:
\begin{eqnarray}\label{groundstate1a}
{\mathbf s}(0,n_1,n_2)&=&\left(\sqrt{1-\tan^2\phi}\;b_0\;{\sf e}^{{\sf i}{\mathbf n}\cdot{\mathbf q}_b},a_0\,{\sf e}^{{\sf i}{\mathbf n}\cdot{\mathbf q}_a}\right)\\
\nonumber
  &=&\left(\sqrt{1-\tan^2\phi}{\sf e}^{{\sf i}\,(n_1-n_2) \frac{2\pi}{3}}, -\tan\phi {\sf e}^{{\sf i}\,n_2 \pi} \right)\\
  \label{groundstate1b}
  &&\\
  \label{groundstate1c}
{\mathbf s}(1,n_1,n_2)&=&\left(\sqrt{1-\tan^2\phi}\;b_1\;{\sf e}^{{\sf i}{\mathbf n}\cdot{\mathbf q}_b},a_1\,{\sf e}^{{\sf i}{\mathbf n}\cdot{\mathbf q}_a}\right)\\
 \label{groundstate1d}
 &=&\left(0,{\sf e}^{{\sf i}\,n_2\,\pi}\right)
   \;,
\end{eqnarray}
taking into account that ${\sf e}^{{\sf i}\,n_2\,\pi}=(-1)^{n_2}$ is real. There occur only $8$ different spin vectors,
see Figure \ref{FIGABC}, {left panel}.
From the explicit form of the ground state configuration (\ref{groundstate1a}-\ref{groundstate1d}),
it is evident that $\phi$ must be restricted to $-\pi/4 \le \phi \le \pi/4$ to make $\sqrt{1-\tan^2\phi}$ real.
These ground states constitute the phases $III$ and $III^\ast$ for  $-\pi/4 \le \phi \le 0$  and  $0 < \phi \le \pi/4$, respectively.
Our investigations indicate that these phases are ``exclusively three-dimensional", i.~e., that no further two- or one-dimensional ground states
exist.

On the other hand, the minimal energy function $E^{(1)}_{\scriptsize min}(\phi)$ according to (\ref{energy1})
seems to hold for the larger interval $-\arctan 3<\phi <\arctan 3$, see Figure \ref{FIGSHE}.
Thus, the task remains to find ground states for $\pi/4 < |\phi| <\arctan 3$ that realize the minimal energy (\ref{energy1})
and constitute the phases $II$ and $II^\ast$.

It turns out that there is a large degeneracy of ground states for the $II$ and $II^\ast$ phases
including two-dimensional as well as three-dimensional states, which
would suggest calling these phases ``classical spin liquids''.
We may expect that thermal or quantum fluctuations will select the
two-dimensional states (``order from disorder") \cite{Vetal80,S82,H89}.
The mentioned degeneracy can be visualized in Figure \ref{FIGHPC}, {right panel,}
where we have chosen $\phi=1.1$,  a Brillouin set ${\mathcal Q}$ with $|{\mathcal Q}|=85$ corresponding to $N_1=N_2=12$
and plotted various functions $\jmath_{\scriptsize min}(\lambda,{\mathbf q})$ for ${\mathbf q}\in{\mathcal Q}$.
All functions meet at the critical point with coordinates $(\hat{\lambda},\jmath_{\scriptsize min}(\hat{\lambda}))$.
Physical ground states can be constructed
by linear combinations of two eigenvectors corresponding to $\jmath_{\scriptsize min}(\hat{\lambda},{\mathbf q}'_a)$ and
$\jmath_{\scriptsize min}(\hat{\lambda},{\mathbf q}'_b)$, respectively, such that $\jmath_{\scriptsize min}({\lambda},{\mathbf q}'_a)$
increases with $\lambda$ and $\jmath_{\scriptsize min}({\lambda},{\mathbf q}'_b)$ decreases, or vice versa.
These ground states are, at most, two-dimensional if both eigenvectors are real and three-dimensional if one is real and one complex.
Note that $\hat{\lambda}$ has always the same value as in (\ref{lambdamax}).

The latter case can be realized by choosing the example
${\mathbf q}'_{a}=(0,0)$ and ${\mathbf q}'_{b}=\left(-\pi ,\frac{\pi }{3}\right)$
The eigenvectors of $\widehat{\mathbbm J}\left({\mathbf q}'_{a},\hat{\lambda}\right)$ and
$\widehat{\mathbbm J}\left({\mathbf q}'_{b},\hat{\lambda}\right)$
corresponding to the eigenvalue (\ref{energy1})
will be
\begin{equation}\label{eigenap}
 {\mathbf a}'=(a'_0,a'_1)=\left(-\frac{\tan \phi }{3},1\right)
 \;,
\end{equation}
and
\begin{equation}\label{eigenbp}
 {\mathbf b}'=(b'_0,b'_1)=\left(
 (-1)^{2/3} \tan \phi ,1\right)
 \;.
\end{equation}
Again, it is straightforward to write the ground state configuration
as the following superposition of ${\mathbf a}'\,{\sf e}^{{\sf i}{\mathbf n}\cdot{\mathbf q}'_{a}}$ and
${\mathbf b}'\,{\sf e}^{{\sf i}{\mathbf n}\cdot{\mathbf q}'_{b}}$:
\begin{eqnarray}\label{groundstate2a}
\gamma'_a&:=&\frac{3}{2} \sqrt{1-\frac{\csc ^2(\phi )}{2}}\;,\\
\label{groundstate2b}
\gamma'_b&:=&\frac{1}{2} \sqrt{\frac{9 \csc ^2(\phi )}{2}-5}\;,\\
\label{groundstate2c}
{\mathbf s}'(0,n_1,n_2)&=&\left(\gamma'_b\;b'_0\;{\sf e}^{{\sf i}{\mathbf n}\cdot{\mathbf q}'_{b}},\gamma'_a\,a'_0\,{\sf e}^{{\sf i}{\mathbf n}\cdot{\mathbf q}'_{a}}\right)\\
\nonumber
  &=&\left(\gamma'_b\, (-1)^{2/3} \tan \phi\,{\sf e}^{{\sf i}\,(-n_1+\frac{n_2}{3}) \pi},-\gamma'_a\,\frac{\tan \phi }{3} \right)\\
  \label{groundstate2d}
    &&\\
  \label{groundstate2e}
{\mathbf s}'(1,n_1,n_2)&=&\left(\gamma'_b\;b'_1\;{\sf e}^{{\sf i}{\mathbf n}\cdot{\mathbf q}'_{b}},\gamma'_a\;a'_1\,{\sf e}^{{\sf i}{\mathbf n}\cdot{\mathbf q}'_{a}}\right)\\
 \label{groundstate2f}
 &=&\left(\gamma'_b\,{\sf e}^{{\sf i}\,(-n_1+\frac{n_2}{3}) \pi},\gamma'_a\right)
   \;.
\end{eqnarray}
The $12$ spin vectors occurring in the ground state ${\mathbf s}'$ for the choice $\phi=\pi/3$ are shown in
Figure \ref{FIGABC}, textcolor{red}{right panel}.
The restriction $\pi/4 \le |\phi| \le \arctan 3$ is necessary to guarantee that $\gamma'_a$ and $\gamma'_b$ will be real.

The case of two-dimensional ground states can be treated by choosing ${\mathbf q}'_{a}=(0,0)$ and ${\mathbf q}'_{b}=(\pi,0)$.
The corresponding eigenvectors read $u=\left(-\frac{\tan (\phi )}{3},1\right)$ and $v=(-\tan (\phi ),1)$.
The procedure analogous to that sketched in Section \ref{sec:OC} leads to a Gram matrix $G=W\,\Delta\,W^\top$ of the ground state where $W=(u,v)$ and
$\Delta$ is the diagonal $2\times 2$-matrix
\begin{equation}\label{Delta1}
\Delta=\left(
\begin{array}{cc}
 \frac{9}{8} \left(1-\cot ^2(\phi )\right) & 0 \\
 0 & \frac{1}{8} \left(9 \cot ^2(\phi )-1\right) \\
\end{array}
\right)
.
\end{equation}
$\Delta$ will be a positively definite matrix for $\frac{\pi}{4}<\phi<\arctan(3)$ and
gives rise to the two-dimensional ground state with values
\begin{equation}\label{groundunitcell}
  \left( \mathbf{s}_0,\mathbf{s}_1\right)=\left( \sqrt{\Delta_{1,1}}\,u,\sqrt{\Delta_{2,2}}\,v\right)^\top
\end{equation}
in the primitive unit cell.
This completes the discussion of the spin liquid phases $II$ and $II^\ast$.

{For those readers who prefer a description of the phase boundaries in terms of the quotient $J_2/J_1= \cot \phi$ we have
provided a Table \ref{T1}  for the case of $J_1>0$, i.~e., anti-ferromagnetic NN coupling.}

\begin{table}
  \centering
  \begin{tabular}{|r|c|c|}\hline
  Phase & Angle & Quotient \\
  \hline\hline
  $III^\ast$ &$0<\phi<\frac{\pi}{4}$& $\infty > \frac{J_2}{J_1}>1$\\
  \hline
  $II^\ast$ &$\frac{\pi}{4}<\phi<\arctan 3$& $1 > \frac{J_2}{J_1}>\frac{1}{3}$\\
  \hline
  $I^\ast$ &$\arctan 3<\phi<\pi$& $\frac{1}{3} > \frac{J_2}{J_1}>-\infty$\\
  \hline
  \end{tabular}
  \caption{{The phase boundaries in terms of $J_2/J_1= \cot \phi$ for $J_1>0$.}}\label{T1}
\end{table}

It remains to investigate the ground states for special values of $\phi$, for example at the phase boundaries.
First we note that for $\phi=0$, i.~e., $J_1=0$ and $J_2=1$ we have a triangular lattice formed of even spins. Its ground states
with minimal energy $-3/2$ according to (\ref{energy1})
are the unique extensions of the well-known local two-dimensional ground state with angle of  $120^\circ$  between adjacent spins,
whereas the odd spins are completely arbitrary.

At the analogous value of $\phi=\pi/2$, i.~e., $J_1=1$ and $J_2=0$ we have a honeycomb lattice with an anti-ferromagnetic NN interaction
and the ground state is one-dimensional with opposite spin directions for even and odd spins in accordance with the one-dimensional phase $I^\ast$ described above.
For the special value of $\phi=\pi/4$ the two interactions are equal, $J_1=J_2=1$, and we have the phase transition between
ground states of phase $II^\ast$ and $III^\ast$. At this point we encounter two-dimensional and one-dimensional ground states that result from the limits of
(\ref{groundstate1a}-\ref{groundstate1d})  and (\ref{groundstate2c}-\ref{groundstate2f}) for $\phi\to\pi/2$. Especially, the one-dimensional ground state
reads ${\mathbf s}(k,n_1,n_2)=(-1)^{n_2+k+1}$.
Moreover, at the value $\phi=\arctan(3)$ we have $\gamma'_b=0$ in  (\ref{groundstate2c}-\ref{groundstate2f})
and the phase transition between ground states of phase $II^\ast$ and $I^\ast$ occurs.

Finally, we consider the special case $\phi=\pi/3$ where the minimal energy of ground states of phase $II^\ast$
assumes the lower Luttinger-Tisza bound, see Figure \ref{FIGSHE}. This can be explained by the occurrence of
another two-dimensional spiral ground state corresponding to the eigenvector $({\sf i},1)$ of
$\widehat{\mathbbm J}(q_1=q_2=2\pi/3,\lambda=0)$ with eigenvalue $-3/2$.

\subsection{Modified square lattice}\label{sec:MSL}

\begin{figure}[htp]
\centering
\includegraphics[width=1.0\linewidth]{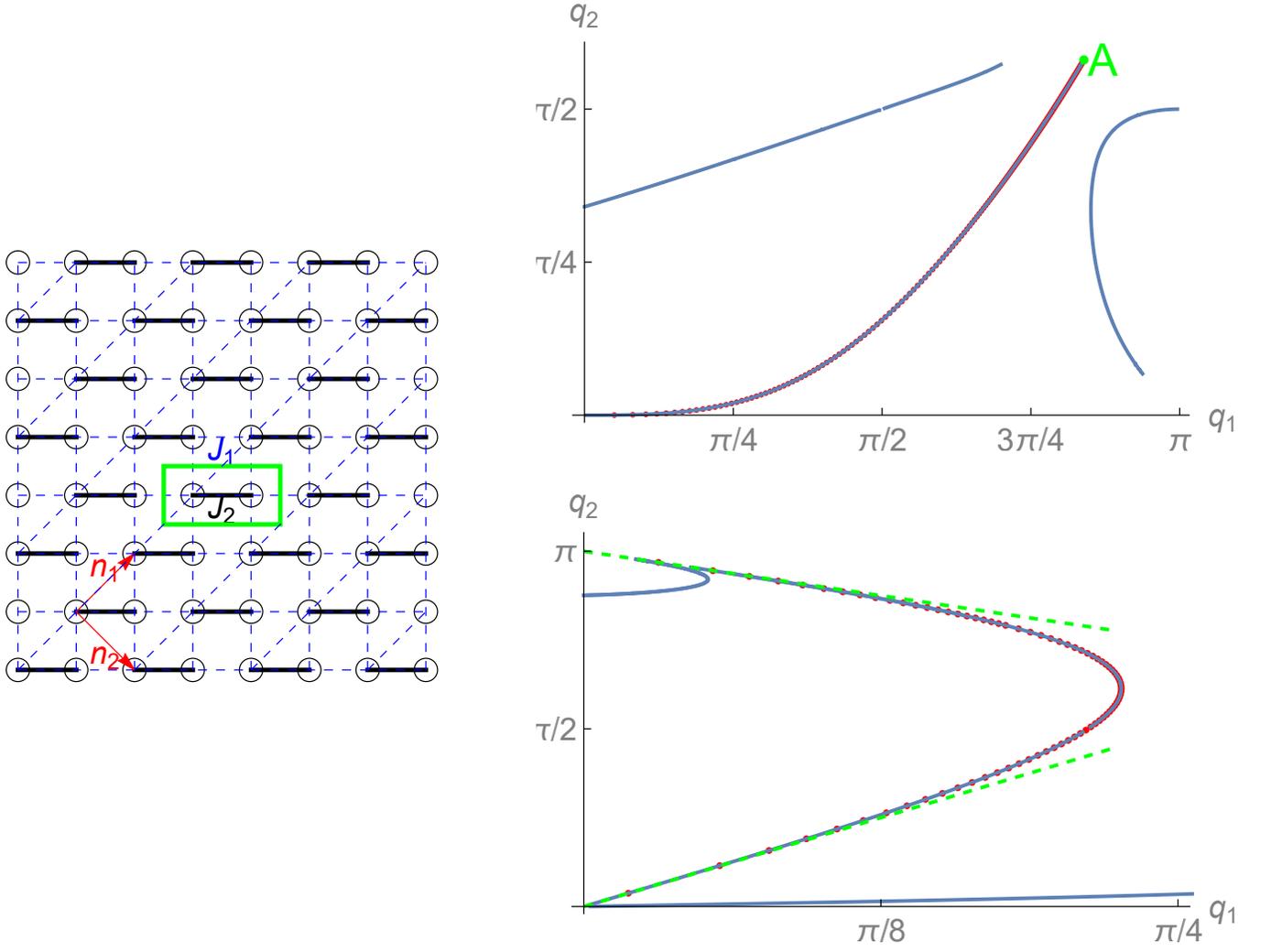}
\caption{
{
Left panel:
Sketch of the modified square lattice. The primitive unit cell (green rectangle) is occupied with two spin sites.
There are two couplings of strength $J_1$ and $J_2$.}\\
{
Right panel:
Plot of various branches of the (blue) curve in ${\mathcal Q}^\infty$
corresponding to the equation (\ref{P1P2}) together with numerical solutions
(red points) representing incommensurable ground states of phase $IV$ (right upper panel) and $V$ (right lower panel).
We have also drawn in the conjectured tangents $q_2=2\,q_1$ and $q_2=\pi-q_1$ (dashed, green lines) for the phase $V$ (right lower panel)
and the (green) endpoint $A$ with coordinates (\ref{coorA}) of the curve corresponding to phase $IV$ (right upper panel).
}
}
\label{FIGGF}
\end{figure}

This example is a modified version of \cite{Ketal00}, where incommensurable ground states were determined analytically.
The modification consists of additional diagonal bonds that create two non-equivalent spin sites with coordinates, say,
${\mathbf s}_0=(0,0),\;{\mathbf s}_1=(1,0)$,  in the primitive unit cell,
see Figure \ref{FIGGF}, {left panel}, and which thus require a generalization of LTLK theory. The lattice is obtained by multiples
of translations of ${\mathbf s}_0$ and ${\mathbf s}_1$ with translation vectors ${\mathbf e}_1=(1,1)$ and ${\mathbf e}_2=(1,-1)$,
see Figure \ref{FIGGF}, {left panel}. Analogously to Section \ref{sec:AH} we write the two coupling constants as
$J_1=\cos\phi$ and $J_2=\sin\phi$ where $-\pi<\phi\le \pi$.
The $2\times 2$-matrix $\widehat{\mathbbm J}({\mathbf q},\lambda)$ assumes the form (\ref{Jq}) with
\begin{eqnarray}
\label{J11}
  J_{11} &=&2 \cos q_1 \cos \phi +\lambda\\
 \nonumber
  J_{12} &=& \overline{J_{21}}=\sin \phi\,+\left({\sf e}^{-{\sf i} q_1}+e^{-{\sf i} q_2}+{\sf e}^{-{\sf i}\left(q_1+q_2\right)}\right) \cos \phi  \\
  && \label{J12}\\
  \label{J22}
  J_{22}&=&-\lambda
  \;.
\end{eqnarray}
Let $p({\mathbf q},\lambda;z)$ denote the characteristic polynomial of $\widehat{\mathbbm J}({\mathbf q},\lambda)$ in the variable $z$.
It turns out that the maximum in (\ref{groundgaugeinfinite}) is always smooth. The corresponding condition
$\frac{\partial p({\mathbf q},\lambda;z)}{\partial \lambda}=0$ leads to
\begin{equation}\label{lambdamsl}
  \hat{\lambda}= -\cos q_1\,\cos\phi
  \;.
\end{equation}

In the sector $-\pi<\phi <-\pi/2$ both coupling constants are negative and the ground state is ferromagnetic, i.~e., all spins have the same direction.
We will see later that this ferromagnetic phase ($I$) can be extended slightly beyond $\phi=-\pi$.
Obviously, the corresponding wave vector is ${\mathbf q}=(0,0)$. The two eigenvectors of $\widehat{\mathbbm J}(0,0,\hat{\lambda})$
are obtained as $(1,1)$ and $(1,-1)$, the first one belonging to the ferromagnetic phase and the second one giving rise to another
one-dimensional phase ($II$) with ${\mathbf q}=(0,0)$ and the two spins in the primitive unit cell being anti-parallel, symbolized by $\uparrow\downarrow$
(N\'{e}el state).
The corresponding minimal energies are obtained as
\begin{eqnarray}
\label{Emsl1}
  E_\text{\scriptsize min}^{(I)} &=& 4\,\cos\phi+\sin\phi\;,\\
 \label{Emsl2}
  E_\text{\scriptsize min}^{(II)} &=& -2\,\cos\phi-\sin\phi
  \;.
\end{eqnarray}
A third one-dimensional phase ($III$) is obtained by setting ${\mathbf q}=(0,\pi)$. It is also of the form $\uparrow\downarrow$ in the primitive
unit cell and has the minimal energy
\begin{equation}\label{Emsl3}
E_\text{\scriptsize min}^{(III)}= \cos\phi - \sin\phi
\;.
\end{equation}
The exact $\phi$-domain of these one-dimensional phases will be determined later.

It turns out that the remaining ground states will be incommensurable.
Partial analytical treatment is possible.
The minimum of ${\jmath}_\text{\scriptsize min}({\mathbf q},\hat{\lambda})$ will be obtained at points ${\mathbf q}\in{\mathcal Q}^\infty$
satisfying
\begin{eqnarray}\label{condmin1}
 p_1&:=&\frac{\partial p({\mathbf q},\hat{\lambda};z)}{\partial q_1}=0\\
 \label{condmin2}
 p_2&:=&\frac{\partial p({\mathbf q},\hat{\lambda};z)}{\partial q_2}=0
 \;.
\end{eqnarray}
$p_2$ is independent of $z$ and (\ref{condmin2}) can be solved for $\phi$ with the relatively simple result
\begin{equation}\label{phisolve2}
 \phi=P_2({\mathbf q}):=\nu\,\pi  +\arctan \left(\frac{\sin\left( \frac{q_1}{2}-q_2\right)}{\sin\left( \frac{q_1}{2}+q_2\right)}\right)
 \;,
\end{equation}
where the integer $\nu\in{\mathbbm Z}$ corresponding to the branch of $\arctan$ has to chosen appropriately.
On the other hand, $p_1$ is linear in $z$ and (\ref{condmin1}) can be solved for $z$. The result can be inserted into $p({\mathbf q},\hat{\lambda};z)$
and yields a function $p_3(\phi,{\mathbf q})$ that vanishes for ground states.
The condition $p_3(\phi,{\mathbf q})=0$ can also be solved for $\phi$, albeit with a more complicated result. It reads
\begin{equation}\label{phisolve1}
 \phi=P_1({\mathbf q}):=\frac{1}{2} \left(\arctan \left(\frac{b c\mp a \sqrt{a^2+b^2-c^2}}{ac \pm b \sqrt{a^2+b^2-c^2}}\right)
 +\nu\pi  \right)
   \;,
\end{equation}
where the sign $\pm$ and $\nu\in{\mathbbm Z}$ have to chosen appropriately and
\begin{eqnarray}
\nonumber
 {a} &{:=}& {-4 \left(-\sin \left(q_1-\frac{3 q_2}{2}\right)+\sin \left(2 q_1-\frac{q_2}{2}\right)+2 \sin
   \left(q_1+\frac{q_2}{2}\right)\right.}\\
   \label{abba}
   &&{\left.
   +\sin \left(\frac{q_2}{2}\right)+\sin \left(q_1+\frac{3   q_2}{2}\right)\right) \cos \left(\frac{q_2}{2}\right) \csc \left(q_1\right)
   }\\
   \nonumber
  {b} & {:=}&
  {-4 \left(\cos \left(q_1\right)-\cos \left(q_2\right) \left(\cos \left(q_2\right)+1\right)+\cos
   \left(q_1+q_2\right)\right.}
   \\
  && {\left.
   +\sin ^2\left(q_2\right) \cot ^2\left(q_1\right)-1\right)
   }\\
   \nonumber
  {c} & \scriptstyle{:=}&
   {
   -2-4 \cos \left(q_1-q_2\right)+4 \cos \left(q_2\right)+2 \cot ^2\left(q_1\right)-\cot
   \left(q_1\right) \csc \left(q_1\right)}\\
   \label{abbc}
   &&
    {
   +\left(\cos \left(3 q_1\right)-2 \cos \left(2 q_1\right)
   \cos \left(2 q_2\right)\right) \csc ^2\left(q_1\right)\;.
   }
\end{eqnarray}
Then the equation
\begin{equation}\label{P1P2}
 P_1({\mathbf q})=P_2({\mathbf q})
\end{equation}
describes a curve in ${\mathcal Q}^\infty$
such that certain branches of it correspond to two families of incommensurable ground states,
denoted by phase $IV$ and $V$, see Figure \ref{FIGGF}, {right panel}.
These branches are determined by numerically solving $p_2=p_3=0$ in the vicinity of finite model solutions.
It is difficult to confirm it by direct calculation, but we conjecture that the branch corresponding to
phase $IV$ has an endpoint $A$ with coordinates
\begin{equation}\label{coorA}
  q^{(A)}_1=\pi -\arccos \left(\frac{7}{8}\right),\;q^{(A)}_2=\frac{1}{2} \left(\pi +\arccos \left(\frac{7}{8}\right)\right)
  \;,
\end{equation}
see Figure \ref{FIGGF}, {right} upper panel. Together with the starting point of this branch at ${\mathbf q}=(0,0)$ with vanishing slope
this implies, according to (\ref{phisolve2}), that the phase $IV$ ranges from $\phi=-\pi/2$ to $\phi=\pi/4$.

Similarly, we conjecture that the branch belonging to phase $V$ has the tangents $q_2=2\,q_1$ and
$q_2=\pi-q_1$ at $q_1=0$, see Figure \ref{FIGGF}, {right} lower panel. In view of (\ref{phisolve2}) this would imply that the incommensurable
ground states of phase $V$ belong to the interval
\begin{eqnarray}\label{interval2a}
&&a \le \phi \le b,\quad \text{where}\\
\label{interval2b}
&& a:= \pi-\arctan 3 \approx 1.89255\quad\text{and } \\
\label{interval2c}
&& b:=\pi-\arctan {\scriptsize \frac{3}{5}}\approx 2.60117
 \;,
\end{eqnarray}
such that the lower limit corresponds to ${\mathbf q}=(0,\pi)$ and the upper one to ${\mathbf q}=(0,0)$.

\begin{figure}[htp]
\centering
\includegraphics[width=0.7\linewidth]{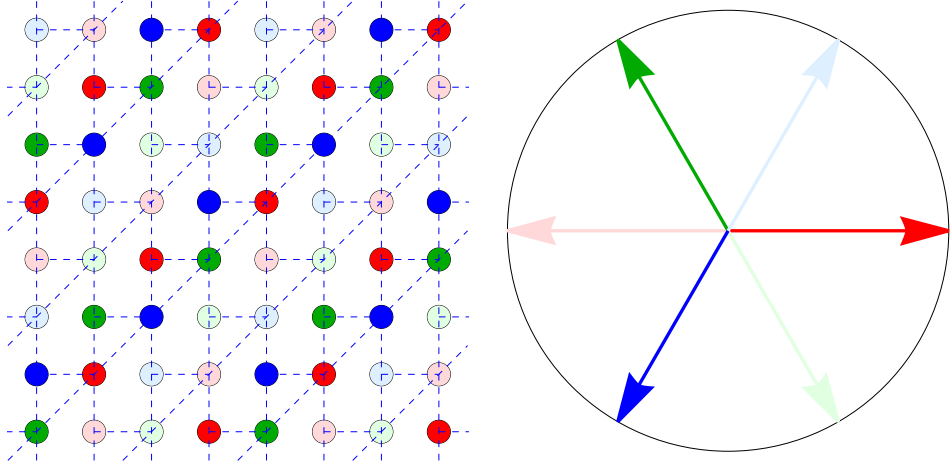}
\caption{Sketch of a ground state of the modified square lattice for  $J_1=1,J_2=0$ ({left} panel) where the spins assume only six different directions
forming a regular hexagon indicated by six different colors according to the {right} panel. This spin lattice can also be viewed
as a system of AF-coupled sawtooth chains. Other ground states may be constructed by rotating the local ground states of the triangles
in one sawtooth chain and accordingly in the other chains. This yields a large degeneracy of ground states at $J_1=1,J_2=0$.
}
\label{FIGGD}
\end{figure}

It is possible that the two families of ground states also contain some states with rational multiples $q_1,q_2$ of $\pi$,
but they will nevertheless be called ``incommensurable" according to convention.
For the boundary of the incommensurable region this is plausible, but also at $\phi=0$, i.~e., $J_1=1,J_2=0$,
there exists a two-dimensional ground state with angles of $120^\circ$ and $180^\circ$ between adjacent spins
that can be constructed elementarily and has $q_1=\frac{2\pi}{3},\;q_2=\frac{\pi}{3}$, see Figure \ref{FIGGD}.

To illustrate the construction of incommensurable ground states we choose the example of $\phi=-\pi/4$,
i.e., $J_1=-J_2=\frac{\sqrt{2}}{2}$.
The corresponding ground state of the infinite lattice can be approximated by a ground state of a finite model such that
$q_1=\frac{19\pi}{24}\approx 2.48709$, $q_2=\pi/2\approx 1.5708$ and $E_{\scriptsize\text{min}}^{(1)}=-2.35473$.
We choose these numbers as initial values for numerically calculating the common root of $p_2=0$ and $p_3=0$.
The result reads $q_1=2.47535$ and $q_2=1.5708$, where remarkably the value for $q_2$ is identical to the initial value.
Inserting these values into $\widehat{J}({\mathbf q},\hat{\lambda})$ we determine its lowest eigenvalue
as $E_{\scriptsize \text{min}}\approx -2.35480$, slightly below $E_{\scriptsize\text{min}}^{(1)}$.
The corresponding eigenvector is $(0.945027 + 0.326993 \,{\sf i},1)$
and can be identified with the incommensurable ground state
\begin{equation}\label{s0s1}
 ({\mathbf s}_0,{\mathbf s}_1)=\left(
 \begin{array}{cc}
  0.945027  & 1 \\
  0.326993  & 0
 \end{array}
 \right)
\end{equation}
in the primitive unit cell,
if we again represent two-dimensional states by complex numbers of absolute value $1$.

If we assume $q_2=\pi/2$ from the outset the equation $p_3=0$ can be solved analytically with the result
\begin{equation}\label{q1analyt}
q_1=\pi -\arctan \left(\sqrt{\frac{1}{2} \left(\sqrt{5}-1\right)}\right)
\;,
\end{equation}
and the corresponding minimal energy
\begin{equation}\label{Eminanalyt}
E_{\scriptsize \text{min}}=-\frac{1}{2} \sqrt{11+5 \sqrt{5}}
\;.
\end{equation}
The corresponding eigenvector has the form
\begin{equation}\label{eigvectoranalyt}
  \left(\frac{1}{\sqrt[4]{-2+\sqrt{5}-2 {\sf i} \sqrt{\sqrt{5}-2}}},1\right)
  \;.
\end{equation}
All these analytical results agree with the above numerical values.

\begin{figure}[htp]
\centering
\includegraphics[width=0.7\linewidth]{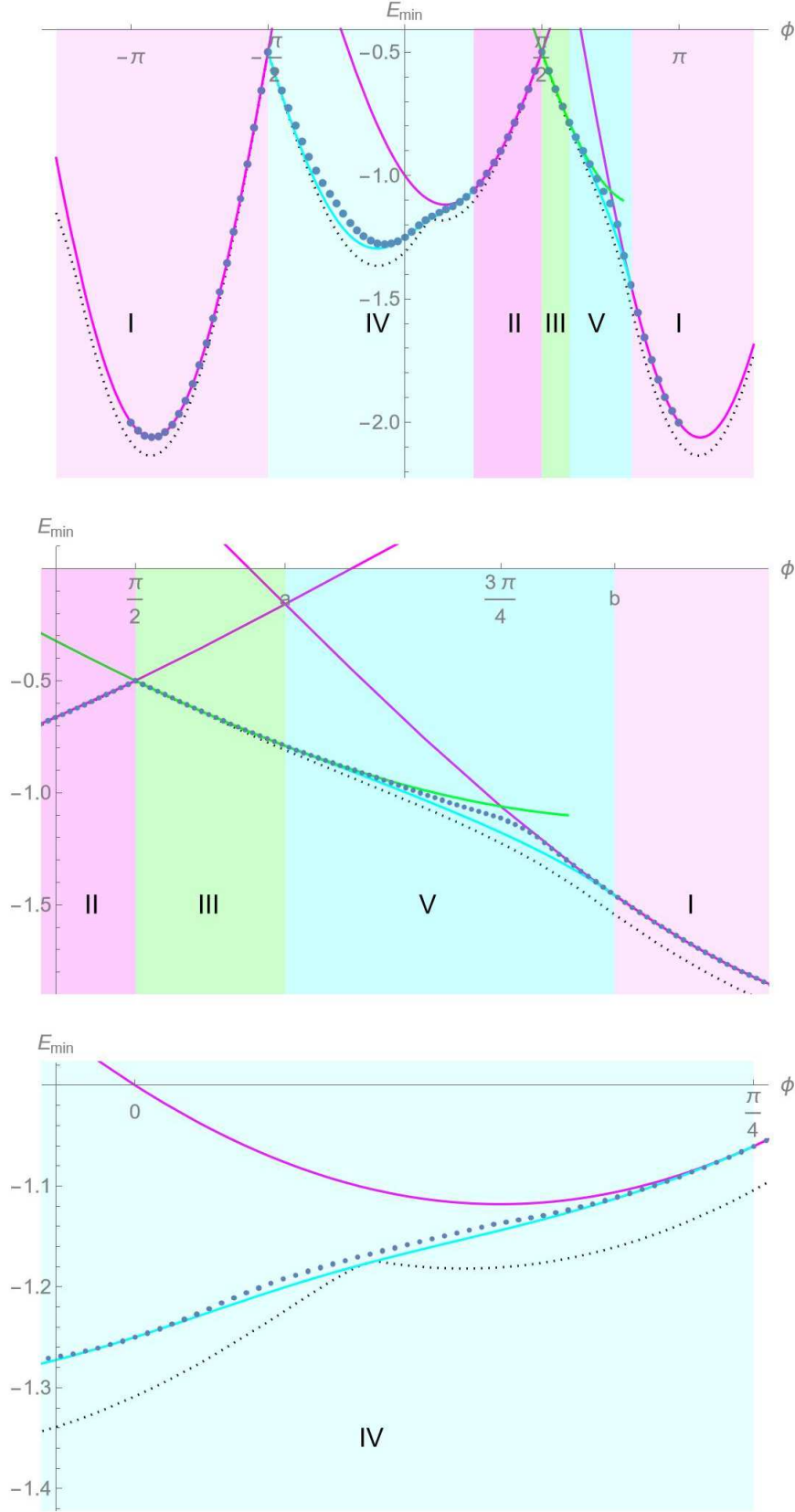}
\caption{Plot of a minimal energy of the $J_1-J_2$ modified square lattice as a function of $\phi$. The black curve represents
the minimal eigenvalues of the $J$-matrix with $\lambda=0$ calculated for a finite model with $M=1152$ spins (lower Luttinger-Tisza bound).
The magenta curves represent the minimal energy of the one-dimensional phases $E^{(I)}_{\scriptsize min}$  and
$E^{(II)}_{\scriptsize min}$ according to (\ref{Emsl1}) and (\ref{Emsl2}),
and the green curve analogously for $ E^{(III)}_{\scriptsize min}$ according to  (\ref{Emsl3}).
The minimal energies of the incommensurable phases $IV$ and $V$ have been calculated semi-analytically and shown by cyan curves.
The blue points represent the numerical Monte Carlo results for the minimal energy
calculated for a finite model with $M=72$ spins.
The upper panel represents the global view, and the middle and the lower one the magnifications of the regions $1.4\le \phi \le 2.9$
and $-0.1\le \phi \le \pi/4$.
The boundaries $a,b$ of the incommensurable phase $V$ are given in (\ref{interval2b},\ref{interval2c}).
All minimal energies are multiplied with a global factor of $1/2$ according to the convention used in large parts of the literature.
We distinguish between five phases ($I - V$) indicated by light colors as explained in the text.
}
\label{FIGSL123}
\end{figure}

When we combine all this information, we get a phase diagram, see Figure \ref{FIGSL123}. Additionally, we have
calculated $E_{\scriptsize \text{min}}(\phi)$ for a finite model with $M=1152$ spins without ground-state gauge.
This would represent a lower Luttinger-Tisza bound only for $M=1152$, but it turns out that it is a lower bound also for the infinite system,
see Figure \ref{FIGSL123}. Numerical Monte Carlo calculations of $E_{\scriptsize \text{min}}(\phi)$ have been performed for an $M=72$ model.
These values coincide with the theoretical $E_{\scriptsize \text{min}}(\phi)$ for the three one-dimensional phases $I,II$ and $III$,
but due to the mismatch of the periodic boundary conditions and incommensurable spiral ground state they lie above
the minimal energy of the two incommensurable phases $IV$ and $V$, except for the above-mentioned ground state at $\phi=0$,
see Figure \ref{FIGSL123}, lower panel.

\section{Summary}\label{sec:SUM}
In this paper, we have presented a kind of combination of the LTLK approach and the Lagrange-variety approach for the ground state problem
of spin lattices. This results in the following recipe: Calculate the Fourier transformed dressed
$J$-matrix $\widehat{J}({\mathbf q},{\boldsymbol \lambda})$ and find the values for the wave vector
${\mathbf q}\in{\mathcal Q}$ and the Lagrange parameters ${\boldsymbol \lambda}\in\Lambda$ where the minimal eigenvalue of the $J$-matrix w.~r.~t.~${\mathbf q}$
assumes its global maximum w.~r.~t.~${\boldsymbol \lambda}$. The value for ${\boldsymbol \lambda}\in\Lambda$
is known to be uniquely determined for each (finite) spin lattice and has been denoted by $\hat{\boldsymbol \lambda}$ (ground state gauge).
This first step can be difficult in general, but we have given two non-trivial examples
(the modified honeycomb lattice and the modified square lattice) where it is feasible.

The second step leads to the construction of ground states as linear combinations of the eigenvectors corresponding to the
${\mathbf q}\in{\mathcal Q}$ and $\hat{\boldsymbol \lambda}\in\Lambda$ found in the first step. This step could lead to spin configurations
of an un-physical dimension exceeding three and thus may fail.
In the case of at most three-dimensional ground states, however, it leads directly to a determination of these
as in the two examples.

One may then ask whether it is possible to obtain {\em all} ground states by this method if the problems mentioned do not arise.
First, one must admit that the present method finds only {\em symmetric} ground states, i.e., those whose components are simultaneously
eigenvectors of the lattice translation operator.
Second, when considering a finite model of an infinite lattice, one cannot be sure, in general, whether a larger model
would yield further ground states realizing the same minimum energy per site or even lower the minimum energy.
The latter could only be ruled out by other arguments, e.~g.,
by showing that the lattice is composed of subunits whose energy is already minimal for a given state.
Those ground states of the infinite lattice that cannot be obtained already as ground states of finite models are called ``incommensurable"
in the literature. We have shown that in the second example, the modified square lattice, our method can be extrapolated in order to obtain also
incommensurable ground states.

\appendix

\section{Proofs}\label{sec:A}
\subsection{Proof of Proposition \ref{P1}}\label{sec:A1}

\begin{enumerate}
  \item
  We obtain
   \begin{eqnarray}
\label{proofP1a}
  \widehat{\mathbbm J}^{ij}({\boldsymbol\lambda},{\mathbf q})
   &\stackrel{(\ref{FourJ})}{=}&
  \sum_{{\boldsymbol \ell}}{\mathbbm J}_{{\mathbf 0}, {\boldsymbol \ell}}^{ij}({\boldsymbol\lambda})\,
  {\sf e}^{{\sf i}\,{\boldsymbol \ell}\cdot{\mathbf q}} \\
  \label{proofP1b}
  &\stackrel{(\ref{symmJ})}{=}&
  \sum_{{\boldsymbol \ell}}{\mathbbm J}_{{\boldsymbol \ell},{\mathbf 0}}^{ji}({\boldsymbol\lambda})\,
  {\sf e}^{{\sf i}\,{\boldsymbol \ell}\cdot{\mathbf q}} \\
  \label{proofP1c}
  &\stackrel{(\ref{latticetrans})}{=}&
  \sum_{{\boldsymbol \ell}}{\mathbbm J}_{{\mathbf 0},-{\boldsymbol \ell}}^{ji}({\boldsymbol\lambda})\,
  {\sf e}^{{\sf i}\,{\boldsymbol \ell}\cdot{\mathbf q}} \\
  \label{proofP1d}
  &=&
   \sum_{{\boldsymbol \ell}}{\mathbbm J}_{{\mathbf 0},{\boldsymbol \ell}}^{ji}({\boldsymbol\lambda})\,
  {\sf e}^{-{\sf i}\,{\boldsymbol \ell}\cdot{\mathbf q}}\\
  \label{proofP1e}
  &=& \overline{\sum_{{\boldsymbol \ell}}{\mathbbm J}_{{\mathbf 0},{\boldsymbol \ell}}^{ji}({\boldsymbol\lambda})\,
  {\sf e}^{{\sf i}\,{\boldsymbol \ell}\cdot{\mathbf q}} }\\
  \label{proofP1f}
   &\stackrel{(\ref{FourJ})}{=}&
   \overline{ \widehat{\mathbbm J}^{ji}({\boldsymbol\lambda},{\mathbf q})}
   \;,
\end{eqnarray}
for all $i,j=1,\ldots,L$, where the overline in (\ref{proofP1e}) and (\ref{proofP1f}) denotes the complex conjugate.
  \item
  This follows from
  \begin{eqnarray}
  \label{proofP1g}
  \widehat{\mathbbm J}^{ij}({\boldsymbol\lambda},{\mathbf q})
   &\stackrel{(\ref{proofP1d})}{=}& \sum_{{\boldsymbol \ell}}{\mathbbm J}_{{\mathbf 0}, {\boldsymbol \ell}}^{ji}({\boldsymbol\lambda})\,
  {\sf e}^{-{\sf i}\,{\boldsymbol \ell}\cdot{\mathbf q}}\\
  \label{proofP1h}
   &\stackrel{(\ref{FourJ})}{=}&\widehat{\mathbbm J}^{ji}({\boldsymbol\lambda},{-\mathbf q})
   \;.
  \end{eqnarray}
  \item Two matrices that are transposes of each other have the same eigenvalues. The second part of the claim follows from
  \begin{eqnarray}
   \label{proofP1i}
    &&  \sum_j\widehat{\mathbbm J}^{ij}({\boldsymbol\lambda},{\mathbf q})\, z_j = \jmath(\boldsymbol\lambda,{\mathbf q})\,z_i \\
     \label{proofP1j}
    &\Leftrightarrow&
    \sum_j\overline{\widehat{\mathbbm J}^{ij}({\boldsymbol\lambda},{\mathbf q})}\,\overline{z_j} = \jmath(\boldsymbol\lambda,{\mathbf q})\,\overline{z_i} \\
     \label{proofP1k}
     &\stackrel{(\ref{proofP1f})}{\Leftrightarrow}&
    \sum_j{\widehat{\mathbbm J}^{ji}({\boldsymbol\lambda},{\mathbf q})}\,\overline{z_j} = \jmath(\boldsymbol\lambda,{\mathbf q})\,\overline{z_i} \\
     \label{proofP1l}
   &\stackrel{(\ref{proofP1h})}{\Leftrightarrow}&
     \sum_j\widehat{\mathbbm J}^{ij}({\boldsymbol\lambda},-{\mathbf q})\,\overline{z_j} = \jmath(\boldsymbol\lambda,{\mathbf q})\,\overline{z_i}
     \;,
  \end{eqnarray}
  \end{enumerate}
  thereby completing the proof of Proposition \ref{P1}.
 \hfill$\Box$\\

\subsection{Proof of Proposition \ref{P2}}\label{sec:A2}
We will call ${\mathbf q}^{(k)}$ of {\em real type} iff $\exp\left({\sf i} {\mathbf q}^{(k)}\cdot{\mathbf n} \right)$ is always real (and hence $\pm 1$)
i.~e., iff ${\mathbf q}^{(k)}$ has only the components $0$ or $\pi$. Otherwise it is called of {\em complex type}.
We have to distinguish three cases.
\begin{enumerate}
  \item All ${\mathbf q}^{(k)}$ are different and of real type. In this case the extension of the local state is given by
  \begin{equation}\label{ext1a}
   {\mathbf s}_{i{\mathbf n}}^{(k)}=
   \sum_\ell \underbrace{\Gamma_{\ell k}}_{\gamma_\ell\,\delta_{\ell k}}
   {\mathbf z}_i^{(\ell)}\underbrace{\exp\left({\sf i} {\mathbf q}^{(k)}\cdot{\mathbf n} \right)}_{\pm 1}
   =\pm\,\gamma_k\,{\mathbf z}_i^{(k)}
   \;,
  \end{equation}
  and
  \begin{equation}\label{sum1}
   \sum_k  \left( {\mathbf s}_{i{\mathbf n}}^{(k)}\right)^2= \sum_k  \left( \gamma_k\,{\mathbf z}_i^{(k)}\right)^2
   =
 \sum_k  \left( {\mathbf s}_{i{\mathbf 0}}^{(k)}\right)^2 =1
 \;.
 \end{equation}
 Here we have use in (\ref{ext1a}) that the linear combination is admissible and hence $\Gamma$ must be diagonal
 and in (\ref{sum1}) that the local configuration in the primitive unit cell consists of unit vectors.

 \item All ${\mathbf q}^{(k)}$ are equal, say,  ${\mathbf q}^{(k)}={\mathbf q}^{(0)}$ and of real type.
 In this case the extension of the local state is given by
 \begin{equation}\label{ext2}
  {\mathbf s}_{i{\mathbf n}}^{(k)}=
   \sum_\ell\Gamma_{\ell k}
   {\mathbf z}_i^{(\ell)}\underbrace{\exp\left({\sf i} {\mathbf q}^{(0)}\cdot{\mathbf n} \right)}_{\pm 1}
   \;,
 \end{equation}
   and
  \begin{equation}\label{sum2}
   \sum_k  \left( {\mathbf s}_{i{\mathbf n}}^{(k)}\right)^2= \sum_k  \left(  \sum_\ell\Gamma_{\ell k}
   {\mathbf z}_i^{(\ell)}\right)^2
   =
 \sum_k  \left( {\mathbf s}_{i{\mathbf 0}}^{(k)}\right)^2 =1
 \;.
 \end{equation}

 \item One  ${\mathbf q}^{(k)}$, say, ${\mathbf q}^{(1)}$ is of real type and the two remaining ones are complex,
 such that ${\mathbf q}^{(3)}=-{\mathbf q}^{(2)}$. In this case $\Gamma$ must be diagonal and the extension is given by
 \begin{eqnarray}\label{ext3a}
   {\mathbf s}_{i{\mathbf n}}^{(1)}&=& \Gamma_{11} \, {\mathbf z}_i^{(1)} \,
   \underbrace{\exp\left({\sf i} {\mathbf q}^{(1)}\cdot{\mathbf n} \right)}_{\pm 1}\;,\\
   \label{ext3b}
    {\mathbf s}_{i{\mathbf n}}^{(2)}&=& \Gamma_{22} \,
    \Re\left({\mathbf z}_i^{(2)}\, \exp\left({\sf i} {\mathbf q}^{(2)}\cdot{\mathbf n} \right) \right)\;,\\
    \label{ext3c}
     {\mathbf s}_{i{\mathbf n}}^{(3)}&=& \Gamma_{22} \,
    \Im\left({\mathbf z}_i^{(2)}\, \exp\left({\sf i} {\mathbf q}^{(2)}\cdot{\mathbf n} \right) \right)
    \;,
 \end{eqnarray}
 and
 \begin{eqnarray}
 \label{sum3a}
    \sum_k  \left( {\mathbf s}_{i{\mathbf n}}^{(k)}\right)^2 &=& \left( \Gamma_{11} \, {\mathbf z}_i^{(1)}\right)^2 \\
    && +\Gamma_{22}^2 \left| {\mathbf z}_i^{(2)}\, \exp\left({\sf i} {\mathbf q}^{(2)}\cdot{\mathbf n} \right) \right|^2 \\
    &=&  \left( \Gamma_{11} \, {\mathbf z}_i^{(1)}\right)^2++\Gamma_{22}^2 \left| {\mathbf z}_i^{(2)} \right|^2 \\
    &=& \sum_k  \left( {\mathbf s}_{i{\mathbf 0}}^{(k)}\right)^2=1
    \;,
  \end{eqnarray}
 \end{enumerate}
 thereby completing the proof of Proposition \ref{P2}.  \hfill$\Box$\\

\end{document}